\documentclass[10pt,twocolumn,final,dvips]{IEEEtran}
\usepackage{amsmath,amssymb,epsfig,color}
\usepackage[breaklinks=true, colorlinks=true, linkcolor=dblue, urlcolor=dblue,
citecolor=dblue, pdfpagemode=None, pdfstartview=FitH]{hyperref}

\newtheorem{theorem}{Theorem}

\newtheorem{corollary}{Corollary}
\definecolor{gray}{cmyk}{.2,0.2,.3,.1}
\definecolor{dred}{cmyk}{0,0.9,0.4,0.3}
\definecolor{dblue}{rgb}{0,0,0.5}
\definecolor{dgreen}{rgb}{0,0.3,0}
\definecolor{dgray}{rgb}{0.3,0.3,0}

\title{A Scalable Protocol for Cooperative Time Synchronization
Using Spatial Averaging
  \thanks{The authors are with the School of Electrical and Computer
  Engineering, Cornell University, Ithaca, NY.  URL:
  {{\tt http://cn.ece.cornell.edu/}}.
  Work supported by the National
  Science Foundation, under awards CCR-0238271 (CAREER), CCR-0330059,
  and ANR-0325556.}}
\author{An-swol Hu \hspace{2cm} Sergio D.\ Servetto}

\begin{document}
\maketitle

\begin{picture}(0,0)
\put(0,55){\tt\small Submitted to the IEEE/ACM Transactions on
  Networking, October 2006.}
\end{picture} 
\vspace{-4mm}

\begin{abstract}
Time synchronization is an important aspect of sensor
network operation. However, it is well known that synchronization
error accumulates over multiple hops. This presents a challenge for
large-scale, multi-hop sensor networks with a large number of nodes
distributed over wide areas. In this work, we present a protocol
that uses spatial averaging to reduce error accumulation in
large-scale networks. We provide an analysis to quantify the
synchronization improvement achieved using spatial averaging and
find that in a basic cooperative network, the skew and offset
variance decrease approximately as $1/\bar{N}$ where $\bar{N}$ is
the number of cooperating nodes.  For general networks, simulation
results and a comparison to basic cooperative network results are
used to illustrate the improvement in synchronization performance.

\end{abstract}

\section{Introduction}

\subsection{Synchronization and the Scalability
Problem} \label{sec:scalability_problem}

Many synchronization techniques have been proposed for synchronizing
sensor networks~\cite{ElsonGE:02,SichitiuV:03,
GaneriwalKW:03,GreunenR:03,MarotiKSL:04}. These techniques all rely
on nodes exchanging packets with timing information.  Using the
exchanged timing information, each node can then estimate clock
offset and maybe clock skew. However, all of these traditional
synchronization techniques suffer from an inherent scalability
problem---synchronization error accumulates over multiple hops.  At
each hop, nodes will estimate synchronization parameters, but the
estimates will have errors.  Therefore, when these erroneous
parameters are used to communicate timing information to the next
hop, errors will further increase.

This accumulation of error over multiple hops poses a problem as
sensor networks are deployed over larger and larger areas. The
number of hops required to communicate across the network increases
and, thus, the synchronization error across the network increases as
well. One possible way to avoid the scalability problem is to use a
few nodes with powerful radios to limit the number of hops required
to communicate timing information across the network.  However, this
technique does not address the fundamental scalability problem of
errors accumulating over multiple hops.

In this work, we consider the use of high density networks to
mitigate the scalability problem.  Recent
developments~\cite{KellyEM:03}, \cite{WarnekLLP:01}, \cite{LiLBH:02}
suggest that future networks may have extremely large numbers of
nodes deployed over wide areas. The question we consider is whether
or not the density of future networks can be used to address
scalability issues that plague existing techniques.

\subsection{Motivation for Cooperation}

In order to reduce the scalability problem, we need to find ways to
reduce the synchronization error at each hop.  There are two primary
ways to accomplish this.  The first is to collect more timing
information. With more timing data, nodes can generally make a
better estimate of clock skew and clock offset.  For example, RBS
and FTSP both let nodes collect many timing data points before
estimating clock skew and clock offset. A timing data point provides
a node with the time at a reference clock at a specific time in its
local time scale. With more data points, synchronization error will
decrease. This idea is essentially doing a \emph{time} average to
estimate clock skew and clock offset. However, this is not
necessarily practical since it would significantly increase the time
to synchronize and the amount of network traffic.

The second primary approach is to improve the quality of the timing
data point.  For example, TPSN and FTSP use MAC layer time stamping
techniques that are more accurate. However, we believe that there is
a fundamentally new technique for improving data point quality that
has not been considered before.  This new idea is to use spatial
averaging to improve data point quality.  In a high density network,
we have a large number of surrounding nodes. Instead of only doing a
time average to estimate the clock skew and clock offset, perhaps we
can also do a \emph{spatial} average to improve these estimates.

\subsection{Approach to Cooperation} \label{sec:approach}

\begin{figure}[!ht]
\centerline{\psfig{file=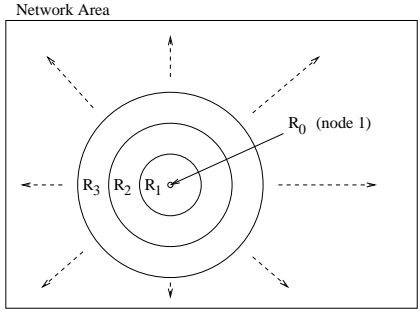,width=5.5cm}}
\caption{\small For increasing $i$, the $R_{i}$ nodes are
progressively farther and farther away from reference node $1$, the
$R_{0}$ node.  Each node in the set $R_{i}$ receives its timing
information from a group of nodes in $R_{i-1}$.}
\label{fig:nodeSets}
\end{figure}

We assume the network is setup such that one node, called node $1$,
has the reference clock that all other nodes want to synchronize to.
Node $1$ will communicate timing information to the nodes in its
broadcast domain, the $R_{2}$ nodes.  The $R_{2}$ nodes will then
communicate timing information to the nodes that are another hop
out, the $R_{3}$ nodes. This process continues until all nodes are
synchronized (Fig.~\ref{fig:nodeSets}).

Each node in $R_{i}$, $i\geq 1$, will use information from the
$R_{i-1}$ nodes to estimate its clock skew and clock offset.  With
these parameters, the node will be able to send a sequence of $m$
pulses that are approximately $d$ seconds apart in the reference
time scale, where $d$ and $m$ are pre-specified by the protocol. All
nodes in the $R_{i}$ set will be attempting to send pulses at the
same time. However, due to synchronization error, the pulses will
only be occurring at approximately the same time. Thus, any node in
the $R_{i+1}$ set of nodes will observe $m$ clusters of pulses
instead of just $m$ individual pulses.

Since each pulse in a cluster represents one $R_{i}$ node's attempt
to transmit at some appropriate time in the reference time scale,
taking the sample mean of the pulse arrival times in each cluster
allows us to average out some of the error made by any one node in
$R_{i}$.  The process of having each node in $R_{i+1}$ use the
sample mean of each cluster as a timing data point is the key to
spatial averaging.  Since each $R_{i+1}$ node can use timing
information from many surrounding neighbors in $R_{i}$, we call our
technique \emph{cooperative time synchronization}.  Using these
timing data points, and some additional information from the $R_{i}$
set, every node in $R_{i+1}$ can estimate its clock skew and offset.
Thus, this process can then repeat to synchronize the $R_{i+2}$
nodes.

The difficulty in studying this problem is that, generally, the $m$
clusters observed by any particular node in $R_{i+1}$ will be
different from the $m$ clusters observed by any other node in
$R_{i+1}$.  This is because the clusters observed by a node will
depend on where this node is located relative to the $R_{i}$ set of
nodes.  Therefore, to study how cooperation can improve
synchronization, we approach the problem in two steps.

\begin{figure}[!ht]
\centerline{\psfig{file=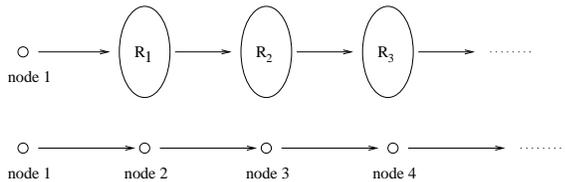,width=7.5cm}}
\caption{\small Top: A basic cooperative network (Type I network)
where timing information is communicated from the $R_{i}$ set of
nodes to the $R_{i+1}$ set. This Type I deployment assumes many
nodes in each $R_{i}$ set. Bottom: Assuming each $R_{i}$ set has
only one node, we have the non-cooperative situation. }
\label{fig:typeIcomparison}
\end{figure}

First, we set up a basic cooperative network (Type I network) that
is a base case for cooperation.  The key assumption in a Type I
network is that all nodes in $R_{i+1}$ are in the broadcast domain
of all nodes in $R_{i}$.  Note that this is a generalization of the
non-cooperative situation where timing information is passed from
one node to the next. Fig.~\ref{fig:typeIcomparison} compares the
basic cooperative network to a non-cooperative network. With the
Type I network we analytically quantify how the variance of the skew
and offset estimates grow with increasing hop number.

Second, we use the theoretical results from the basic cooperative
network to understand the behavior of a general Type II network
where nodes are uniformly distributed over a circular area.
Simulation results are used to illustrate that increasing network
density improves synchronization performance.

\subsection{Other Related Work} \label{sec:relatedwork}

The traditional synchronization techniques describe
in~\cite{ElsonGE:02,SichitiuV:03,
GaneriwalKW:03,GreunenR:03,MarotiKSL:04} all operate fundamentally
on the idea of communicating timing information from one set of
nodes to the next.  One other approach to synchronization that has
recently received much attention is to apply mathematical models of
natural phenomena to engineered networks.  A model for the emergence
of synchrony in pulse-coupled oscillators was developed
in~\cite{MirolloS:90} for a fully-connected group of identical
oscillators. In~\cite{LucarelliW:04}, this convergence to synchrony
result was extended to networks that were not fully connected.

The convergence result is clearly desirable for synchronization in
networks and in~\cite{HongS:04} theoretical and simulation results
suggested that such a technique could be adapted to communication
and sensor networks.  Experimental validation for the ideas
of~\cite{MirolloS:90} was obtained in~\cite{Werner-AllenTPWN:05}
where the authors implemented the Reachback Firefly Algorithm (RFA)
on TinyOS-based motes.

The problem with these emergent synchronization results is that the
fundamental theory assumes all nodes have nearly the same firing
period. Results from~\cite{HongS:04} and~\cite{Werner-AllenTPWN:05}
show that the convergence results may hold when nodes have
approximately the same firing period, but the authors
of~\cite{Werner-AllenTPWN:05} explain that clock skew will degrade
synchronization performance. Since we are not aware of any results
that provide an extension to deal with networks of nodes with
arbitrary firing periods, our work focuses on synchronization
algorithms that explicitly estimate clock skew.

\subsection{Contributions and Paper Organization}

In this paper, we propose a protocol for time synchronization that
uses spatial averaging to improve synchronization performance. In
this work we make the following analysis:
\begin{enumerate}
\item Mathematically quantify the synchronization error for the Type
I basic cooperative network.
\item Through simulations, show that increasing node density can
decrease synchronization error in general networks.
\end{enumerate}

The results show that if each node can hear a large number of
neighboring nodes, then nodes can cooperatively generate signals
that are less noisy and allow for better synchronization performance
over multiple hops.  The fact that more cooperating nodes yields
better performance means that there exists a new trade-off between
network density and synchronization performance where more nodes
provide better synchronization.  Even though it is possible to
achieve better synchronization performance by introducing nodes with
powerful radios to synchronize a large-scale network, cooperative
time synchronization is an effective alternative technique to
reducing synchronization error across the network without requiring
special nodes.

The remainder of the paper is organized as follows.  In
Section~\ref{sec:systemModel} we set up the general network
assumptions and present the synchronization protocol in
Section~\ref{sec:synchronizationProtocol}. The analysis and
simulations of the protocol for a basic cooperative network are
presented in Section~\ref{sec:typeI} while a study of cooperation in
general networks is carried out in Section~\ref{sec:typeII}. We make
concluding remarks in Section~\ref{sec:conclusion}.

\section{System Model} \label{sec:systemModel}

\subsection{Clock Model}

The behavior of each node $i$ is governed by a clock $c_{i}$ that
counts up from $0$. The introduction of $c_{i}$ is important since
it provides a consistent timescale for node $i$. This is the
node's local time scale and in synchronization the node tries to
estimate how its local clock is related to the reference clock.

We assume that node $1$ contains the reference clock and every
node in the network is to be synchronized to this clock.  The
clock of node $1$, $c_{1}$, will be defined as $c_{1}(t)=t$ where
$t \in [0,\infty)$. We now define the clock of any other arbitrary
node $i$, $c_{i}$, as
\begin{equation} \label{eq:clock}
c_{i}(t)=\alpha_{i}(t-\bar{\Delta}_{i})+\Psi_{i}(t),
\end{equation}
where
\begin{itemize}
\item $\bar{\Delta}_{i}$ is an unknown offset between the start
  times of $c_{i}$ and $c_{1}$.
\item $\alpha_{i}>0$ is an unknown constant for each $i$. \item
$\Psi_{i}(t)$ is a stochastic process modelling random timing
jitter. $\Psi_{i}(t)$ is a zero mean Gaussian process with
independent and identically distributed Gaussian samples with mean
zero and variance $\sigma^2$, i.e. $\mathcal{N}(0,\sigma^2)$. We
assume $\sigma^{2}<\infty$ and note that $\sigma^{2}$ is defined
in terms of the clock of node $i$.
\end{itemize}
Note that this linear relationship is valid for short periods of
time since we do not explicitly model clock drift.

\subsection{Transmission Model}

Each node $i$ in the network can transmit short pulses $p(t)$ for
time synchronization.  These are short duration pulses, i.e. ultra
wideband pulses, and for our purposes we consider them to be delta
functions $\delta(t)$.  The particular choice of $p(t)$ is not
important.  For the purposes of studying cooperative time
synchronization, we assume a node receiving the pulse can uniquely
determine a pulse arrival time, pulses sent from different nodes do
not overlap, and a node seeing multiple pulses can identify the
different pulse arrival times. Note that only minor modifications of
the protocol are needed to accommodate other types of pulse
shapes~\cite{HuS:05}.

We assume that each node has a transmission range of $R$.  This
means that a node $j$ must be within a distance $R$ from a
transmitting node $i$ in order to hear pulses from node $i$. Note
that the assumption of a circular transmission region is made only
to simplify the illustration of spatial averaging.  The
synchronization protocol proposed in
Section~\ref{sec:synchronizationProtocol} does not require this
assumption and most of the results in this work will hold under more
realistic conditions~\cite{GanesanKWCEW:02,ZhaoG:03}. Since we are
dealing with sensor nodes who have short transmission distances, we
further assume that propagation delay is negligible. We make this
assumption since from~\cite{MarotiKSL:04} we know that propagation
delay is less than $1\mu s$ for distances up to $300$ meters. Some
results on cooperation and propagation delay are available
in~\cite{HuS:05a}.

\section{Synchronization Protocol}
\label{sec:synchronizationProtocol}

To start synchronization, the reference node, node $1$, will send a
sequence of $m$ pulses that are $d$ seconds apart.  Since we assume
the nodes have impulse radio transmitters, each pulse is extremely
narrow in time. The values of $d$ and $m$ are parameters of the
protocol that are established before deploying the network so the
values are known by all nodes in the network. Therefore, in the time
scale of node $1$ the pulses are transmitted at times $\tau_{0},
\tau_{0}+d,\ldots,\tau_{0}+d(m-1)$, where $\tau_{0}$ is the time at
which the synchronization process started. Let node $1$ be the only
element of the $R_{0}$ nodes.

\begin{figure}[!ht]
\centerline{\psfig{file=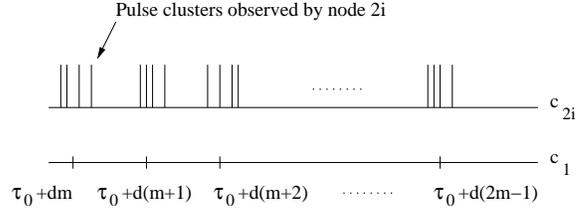,width=7.5cm}}
\caption{\small A node $2i$ in $R_{2}$ has clock $c_{2i}$. This node
will see clusters of pulse arrivals that are transmitted from a
group of nodes in the set $R_{1}$. These clusters arrive at node
$2i$ around times
$\tau_{0}+dm,\tau_{0}+d(m+1),\ldots,\tau_{0}+d(2m-1)$ in the time
scale of node $1$, $c_{1}$.} \label{fig:PulseClusters}
\end{figure}

The nodes that are in the broadcast domain of node $1$ will hear
this sequence of $m$ pulses.  We call these nodes the $R_{1}$ nodes
and each node $i\in R_{1}$, $i\geq 1$, will be denoted by node $1i$.
The vector of pulse arrival times observed by node $1i$ will be
denoted $\mathbf{Y}_{1i}$. Each node $1i$ will be able to estimate
its clock skew since it knows that node $1$ transmitted these pulses
$d$ seconds apart. Each node $1i$ will also predict, in its own time
scale, when times
$\tau_{0}+dm,\tau_{0}+d(m+1),\ldots,\tau_{0}+d(2m-1)$ will occur in
the time scale of node $1$ and transmit $m$ pulses, one at each
predicted time. This means that each node $1i$ will transmit a pulse
approximately at times $\tau_{0}+dm, \tau_{0}+d(m+1),\ldots,
\tau_{0}+d(2m-1)$ in the time scale of node $1$. When the $R_{1}$
nodes each transmit their sequence of $m$ pulses, the nodes that can
hear a subset of the $R_{1}$ nodes, the $R_{2}$ nodes, will observe
clusters of pulses around the times $\tau_{0}+dm,
\tau_{0}+d(m+1),\ldots, \tau_{0}+d(2m-1)$ since each node $2i$ can
hear many $R_{1}$ nodes (Fig.~\ref{fig:PulseClusters}). In fact, we
make sure each node $2i$ can hear a cluster by requiring the node to
observe at least $\bar{N}$ pulses in each cluster.  If a node $2i$
sees less than $\bar{N}$ pulses in a cluster, then it will not make
observations.  Each node $2i$, a node $i\in R_{2}$, will note the
arrival time of each pulse in the $k$th cluster, $k=1,\ldots,m$, and
take the sample mean of these times to be its $k$th observation.
Node $2i$'s vector of observations will be denoted as
$\mathbf{Y}_{2i}$. Using these $m$ observations, any node $2i$ will
be able to estimate its clock skew since it knows that these
observations should be occurring $d$ seconds apart.  As well, it
will be able to predict in its local time scale when times
$\tau_{0}+d(2m), \tau_{0}+d(2m+1),\ldots,\tau_{0}+d(3m-1)$ will
occur in the time scale of the reference time.  Node $2i$ will then
transmit a pulse at each of those predicted times.  This processes
will repeat until all nodes in the network have an estimate of their
clock skew.  Notice that the $R_{1}$ nodes are not required to
observe $\bar{N}$ pulses in each cluster since they will always only
receive a sequence of $m$ pulses from node $1$.  Node $1$ can simply
broadcast a special packet to its surrounding nodes to identify the
$R_{1}$ nodes. An illustration of the process can be found in
Fig.~\ref{fig:nodeSets} and note that nodes will remain silent for
the remainder of the synchronization process after transmitting
their $m$ pulses. The cooperation occurs when a node $ki$ in
$R_{k}$, $k>1$, can take a sample mean of a cluster of pulse
arrivals.

To obtain the clock offset, the $R_{k}$ nodes will broadcast a
packet of information to the $R_{k+1}$ nodes, $k\geq 0$.  This
packet will contain the value of $\tau_{0}$ and a number $q$
denoting the number of hops out from node $1$.  For example, node
$1$ will transmit the value of $\tau_{0}$ and $q=0$ to the $R_{1}$
nodes.  The $R_{1}$ nodes will then send $\tau_{0}$ and $q=1$ to the
$R_{2}$ nodes.  In general, the $R_{k}$ nodes will send $\tau_{0}$
and $q=k$ to the $R_{k+1}$ nodes.  Any node $ki$ will then know that
its first observation approximately occurred at time $\tau_{0}+dmq$
in the time scale of the reference time, where the value of $q$ is
the one received from set $R_{k-1}$.

We now describe how any node $ki$ can estimate its clock skew,
clock offset, and its $m$ pulse transmission times. From
(\ref{eq:clock}), we know that there is a linear relationship
between the reference clock $c_{1}$ and the clock of node $ki$,
$c_{ki}$.  Node $ki$ will have a set of $m$ observations denoted
by the $m\times 1$ vector $\mathbf{Y}_{ki}$, where the elements of
the vector are ordered from the earliest observation time to the
latest observation time. Node $ki$ will estimate its clock skew as
\begin{equation} \label{eq:alphahat}
\hat{\alpha}_{ki} =
\bar{\mathbf{C}}(\mathbf{H}^{T}\mathbf{H})^{-1}\mathbf{H}^{T}\mathbf{Y}_{ki}
\end{equation}
and clock offset as
\begin{equation} \label{eq:Deltahat}
\hat{\Delta}_{ki} =
\tilde{\mathbf{C}}(\mathbf{H}^{T}\mathbf{H})^{-1}\mathbf{H}^{T}\mathbf{Y}_{ki}-(\tau_{0}+dm(k-1)),
\end{equation}
where $\bar{\mathbf{C}}=[0 \quad 1]$, $\tilde{\mathbf{C}}=[1 \quad
0]$ and
\begin{equation}  \label{eq:H}
{\mathbf H} = \left[ \begin{array}{ccccc}
1 & 1 & 1 & \ldots & 1\\
0 & d & 2d & \ldots & (m-1)d
\end{array} \right]^T
\end{equation}
Note that in the calculation of the clock offset
$\hat{\Delta}_{ki}$, the term $\tau_{0}+dm(k-1)$ is the time in
the time scale of $c_{1}$ that node $ki$ should receive its first
pulse.  Node $ki$ has used the $\tau_{0}$ and $q=k-1$ parameters
sent to it from the $R_{k-1}$ nodes.  Node $ki$ will also estimate
its own $m$ pulse transmission times using
\begin{equation} \label{eq:pulsetimeestimate}
X_{ki}(l) =
\mathbf{C}_{l}(\mathbf{H}^{T}\mathbf{H})^{-1}\mathbf{H}^{T}\mathbf{Y}_{ki},
\end{equation}
where $\mathbf{C}_{l} = [1 \quad d(m+l)]$, for $l=0,1,\ldots,m-1$.
$X_{ki}(l)$ is the transmission time of node $ki$'s $(l+1)$th pulse.
A pseudo-code description is given in Table~\ref{tab:protocol}. Note
that the protocol described above is a completely new approach to
the asymptotic spatial averaging ideas we studied in~\cite{HuS:03b}.

\begin{table}[!h]
\caption{\rm The synchronization protocol for each node $ki$, $k >
1$.}
\begin{center}
\fbox{\tt\begin{minipage}{70truemm}\small
\begin{tabbing}
\hspace{2mm} \= {\normalsize Cooperative Time Sync}
  \hspace{1cm} \\ \\
  \> wa\=it for pulse arrivals, at least $\bar{N}$ \\
  \> \>per cluster; \\
  \> while\mbox{ }(number of arrival clusters < $m$) \{ \\
  \> \> record arrival time of all pulses; \\
  \> \> listen for packet with $\tau_{0}$ and $q$ values; \\
  \> \}; \\
  \> for each (pulse arrival cluster $j$) \{\\
  \> \> $\mathbf{Y}_{ki}[j]=$ sample mean of cluster; \\
  \> \}; \\
  \> skew $\hat{\alpha}_{ki} =
\bar{\mathbf{C}}(\mathbf{H}^{T}\mathbf{H})^{-1}\mathbf{H}^{T}\mathbf{Y}_{ki}$;
\\
\> offset $\hat{\Delta}_{ki} =
\tilde{\mathbf{C}}(\mathbf{H}^{T}\mathbf{H})^{-1}\mathbf{H}^{T}\mathbf{Y}_{ki}-(\tau_{0}+dmq)$
\\
\> for (l from $0$ to $m-1$) \{ \\
\> \> transmission time $X_{ki}(l) =
\mathbf{C}_{l}(\mathbf{H}^{T}\mathbf{H})^{-1}\mathbf{H}^{T}\mathbf{Y}_{ki}$;
\\
\> \> transmit pulse at $X_{ki}(l)$; \\
\> \}; \\
\> while (transmitting pulses) \{ \\
\> \> send a packet with values $\tau_{0}$ and $q+1$; \\
\> \}; \\
\end{tabbing}
\end{minipage}}
\label{tab:protocol}
\end{center}
\end{table}

\section{Type I: Basic Cooperative Networks} \label{sec:typeI}

\subsection{Network Setup}

The most basic and fundamental deployment of nodes that effectively
employs cooperative time synchronization is the case where
\emph{all} $\bar{N}$ nodes at any given hop contribute to the
signals observed at the next hop.  This Type I deployment is
illustrated in the top of Fig.~\ref{fig:typeIcomparison} where each
set of nodes $R_{i}$, $i\geq 1$, have $\bar{N}$ nodes.  We see that
every node in $R_{i}$ is in the broadcast domain of every node in
$R_{i-1}$.

\subsection{Analysis} \label{sec:typeIanalysis}

Due to the scalability problem, we would expect synchronization
error to grow as timing information from node $1$ (the $R_{0}$ node)
is communicated to a node in the $R_{k}$ set of nodes, $k>0$.
Therefore, it is of particular interest to quantify how the variance
of the skew and offset estimates change as a function of the hop
number $k$.  Looking at the structure of the basic cooperative
network in Fig.~\ref{fig:typeIcomparison}, we notice that the skew
and offset estimates at a node $ki$ must be dependent only on the
estimates made by the nodes in $R_{k-1}$ since all the information
at $R_{k}$ comes from the $R_{k-1}$ set of nodes.  Therefore, to
understand the synchronization error growth over multiple hops, we
need the distribution of the estimates made by the nodes in $R_{k}$
as a function of the distribution of the estimates made by the nodes
in $R_{k-1}$.  Theorem~\ref{theorem:main}, below, provides us with
this characterization.

In the statement of the theorem we use $e_{l}$ to be the column
vector of all zeros except for a one in the $l$th position and
\begin{eqnarray} \label{eq:barH}
\bar{{\mathbf H}} = \left[ \begin{array}{cccc} {\mathbf H} & 0 &
\ldots & 0 \\
0 & {\mathbf H} & \ldots & 0 \\
\vdots & \vdots & \ddots & \vdots \\
0 & 0 & \ldots & {\mathbf H}
\end{array} \right],
\end{eqnarray}
where ${\mathbf H}$ is from (\ref{eq:H}).  Also, $\alpha_{ki}$ and
$\bar{\Delta}_{ki}$ are the clock model parameters from
(\ref{eq:clock}) for node $ki$.

\begin{theorem} \label{theorem:main}
Assume a Type I basic cooperative network.

(1) Given the distribution of the $2\bar{N}\times 1$ vector of
estimates made by the $R_{k-1}$ nodes,
\begin{equation*}
\hat{\bar{{\mathbf \theta}}}_{k-1} \sim {\mathcal
N}(\bar{\mu}_{k-1}, \bar{\Sigma}_{k-1}),
\end{equation*}
the distribution of the $2\bar{N}\times 1$ vector of estimates made
by the $R_{k}$ nodes,
\begin{equation*}
\hat{\bar{{\mathbf \theta}}}_{k} \sim {\mathcal N}(\bar{\mu}_{k},
\bar{\Sigma}_{k}),
\end{equation*}
is found as follows: $\hat{\bar{{\mathbf \theta}}}_{k}$ has mean
vector
\begin{eqnarray} \label{eq:muk}
\bar{\mu}_{k} &=& E(\hat{\bar{{\mathbf \theta}}}_{k}) \nonumber \\
&=& E(\mathbf{A}_{k}\hat{\bar{\mathbf{\theta}}}_{k-1} +
\mathbf{B}_{k}) \nonumber \\
&=& \mathbf{A}_{k}\bar{\mu}_{k-1} +
\mathbf{B}_{k} \nonumber \\
&=& \left[ \begin{array}{c}
\alpha_{k1}(\tau_{0}+(k-1)md-\bar{\Delta}_{k1}) \\
\alpha_{k1} \\
\vdots \\
\alpha_{k\bar{N}}(\tau_{0}+(k-1)md-\bar{\Delta}_{k\bar{N}}) \\
\alpha_{k\bar{N}} \\
\end{array} \right]
\end{eqnarray}
and covariance matrix
\begin{equation} \label{eq:sigmak}
\bar{\Sigma}_{k} = \textrm{Cov}(\hat{\bar{{\mathbf \theta}}}_{k}) =
\Sigma_{m_{k}}+\mathbf{A}_{k}\bar{\Sigma}_{k-1}\mathbf{A}_{k}^{T}
\end{equation}
for
\begin{eqnarray*}
\Sigma_{m_{k}} & = &
(\bar{\mathbf{H}}^{T}\bar{\mathbf{H}})^{-1}\bar{\mathbf{H}}^{T}\Sigma_{\bar{{\mathbf
W}}_{k}}((\bar{\mathbf{H}}^{T}\bar{\mathbf{H}})^{-1}\bar{\mathbf{H}}^{T})^{T}
\end{eqnarray*}
\begin{eqnarray*}
\Sigma_{\bar{{\mathbf W}}_{k}} = {\mathbf
Q}_{k}\Sigma_{\tilde{\Psi}_{k-1}}{\mathbf Q}_{k}^{T} + \sigma^2
{\mathbf I}_{\bar{N}m}
\end{eqnarray*}
\begin{displaymath}
{\mathbf Q}_{k} = \left[ \begin{array}{c} \alpha_{k1}{\mathbf I}_{m} \\
\vdots \\
\alpha_{k\bar{N}}{\mathbf I}_{m}
\end{array} \right]
\end{displaymath}
\begin{displaymath}
\Sigma_{\tilde{\Psi}_{k-1}} =
\frac{\sigma^{2}}{\bar{N}^{2}}\sum_{i=1}^{\bar{N}}\frac{1}{\alpha_{(k-1)i}^{2}}{\mathbf
I}_{m}
\end{displaymath}
where
\begin{eqnarray*}
\mathbf{A}_{k} = \mathbf{D}_{k}\left[
\begin{array}{ccccc} \frac{1}{\alpha_{(k-1)1}} &
\frac{dm}{\alpha_{(k-1)1}} & \ldots & 0
& 0 \\
0 & \frac{1}{\alpha_{(k-1)1}} & \ldots & 0 & 0 \\
\vdots & \vdots & \ddots & \vdots & \vdots \\
0 & 0 & \ldots & \frac{1}{\alpha_{(k-1)\bar{N}}} &
\frac{dm}{\alpha_{(k-1)\bar{N}}} \\
0 & 0 & \ldots & 0 & \frac{1}{\alpha_{(k-1)\bar{N}}}
\end{array}\right]
\end{eqnarray*}
\begin{eqnarray*}
\mathbf{D}_{k} = \frac{1}{\bar{N}}\left[
\begin{array}{ccccc} \alpha_{k1} & 0  & \ldots &
\alpha_{k1}
& 0 \\
0 & \alpha_{k1} & \ldots & 0 & \alpha_{k1} \\
\vdots & \vdots & \ddots & \vdots & \vdots \\
\alpha_{k\bar{N}} & 0 & \ldots & \alpha_{k\bar{N}}
& 0 \\
0 & \alpha_{k\bar{N}} & \ldots & 0 & \alpha_{k\bar{N}}
\end{array}\right]
\end{eqnarray*}
\begin{eqnarray*}
\mathbf{B}_{k} = \mathbf{D}_{k}\left[ \begin{array}{c}
\bar{\Delta}_{(k-1)1} \\ 0 \\ \vdots \\
\bar{\Delta}_{(k-1)\bar{N}} \\ 0
\end{array}\right] - \left[ \begin{array}{c}
\alpha_{k1}\bar{\Delta}_{k1} \\ 0 \\ \vdots \\
\alpha_{k\bar{N}}\bar{\Delta}_{k\bar{N}} \\ 0
\end{array}\right].
\end{eqnarray*}
The initial conditions are
\begin{eqnarray*}
\bar{\mu}_{1} &=& \left[ \begin{array}{c}
\alpha_{11}(\tau_{0}-\bar{\Delta}_{11}) \\
\alpha_{11} \\
\vdots \\
\alpha_{1\bar{N}}(\tau_{0}-\bar{\Delta}_{1\bar{N}}) \\
\alpha_{1\bar{N}} \\
\end{array} \right]
\end{eqnarray*}
and
$\bar{\Sigma}_{1} = \sigma^2(\bar{{\mathbf H}}^{T}\bar{{\mathbf
H}})^{-1}$.

(2) The skew estimate and offset estimate for node $ki$ can be found
as
\begin{equation} \label{eq:alphahatki}
\hat{\alpha}_{ki} = e_{2(i-1)+2}^{T}\hat{\bar{{\mathbf \theta}}}_{k}
\end{equation}
\begin{equation} \label{eq:Deltahatki}
\hat{\Delta}_{ki} = e_{2(i-1)+1}^{T}\hat{\bar{{\mathbf \theta}}}_{k}
- (\tau_{0}+dm(k-1))
\end{equation}
for $i=1,2,\ldots,\bar{N}$ and $k\geq 1$. \qquad $\bigtriangleup$
\end{theorem}

The proof of Theorem~\ref{theorem:main} is found in the appendix.
Since the distribution of $\hat{\bar{{\mathbf \theta}}}_{k}$ is
available, the distribution of $\hat{\alpha}_{ki}$ and
$\hat{\Delta}_{ki}$ can be found.  In fact, the variance of
$\hat{\alpha}_{ki}$ can be found in element ($2(i-1)+2$, $2(i-1)+2$)
of $\bar{\Sigma}_{k}$ in (\ref{eq:sigmak}) and the variance of
$\hat{\Delta}_{ki}$ can be found in element ($2(i-1)+1$,
$2(i-1)+1$).  The mean of $\hat{\alpha}_{ki}$ is the $(2(i-1)+2)$th
element of $\bar{\mu}_{k}$ in (\ref{eq:muk}) and the mean of
$\hat{\Delta}_{ki}$ can be found from the $(2(i-1)+1)$th element
shifted by $\tau_{0}+dm(k-1)$.

From the statement of Theorem~\ref{theorem:main}, we see that the
distribution of the estimates made by the $R_{k}$ nodes,
$\hat{\bar{{\mathbf \theta}}}_{k}$, is completely determined from
the distribution of $\hat{\bar{{\mathbf \theta}}}_{k-1}$.  This
recursive nature comes from the fact that the parameters estimated
by the $R_{k}$ nodes is only dependent on the estimates made by the
$R_{k-1}$ nodes.  The relationship between $\hat{\bar{{\mathbf
\theta}}}_{k-1}$ and $\hat{\bar{{\mathbf \theta}}}_{k}$ can be
intuitively understood in two steps. First, $\hat{\bar{{\mathbf
\theta}}}_{k-1}$ is the vector of synchronization parameters
estimated by the nodes in $R_{k-1}$. Therefore, these estimates will
establish the synchronization parameters for the $R_{k}$ nodes since
the $R_{k-1}$ nodes communicate timing information to the $R_{k}$
nodes.  The synchronization parameters for $R_{k}$ are found as
\begin{displaymath}
\bar{\mathbf{\theta}}_{k} =
\mathbf{A}_{k}\hat{\bar{\mathbf{\theta}}}_{k-1} + \mathbf{B}_{k}.
\end{displaymath}
Second, the $R_{k}$ nodes will use the timing information from the
$R_{k-1}$ nodes to make an unbiased estimate of the parameters
$\bar{\mathbf{\theta}}_{k}$, which gives us $\hat{\bar{{\mathbf
\theta}}}_{k}$.

Since any node $ki$'s skew and offset estimates are found as affine
transforms of $\hat{\bar{\theta}}_{k}$ in (\ref{eq:alphahatki}) and
(\ref{eq:Deltahatki}), respectively, we see that any estimation
errors made by the $R_{k-1}$ nodes will be propagated to the $R_{k}$
nodes' estimates of clock skew and clock offset.  However, the
intuitive understanding of cooperative time synchronization comes
from realizing that the matrix $\mathbf{A}_{k}$ takes an ``average''
over $\hat{\bar{\mathbf{\theta}}}_{k-1}$ thus mitigating the errors
made by any particular node $(k-1)i$.  As a result, the
synchronization parameters communicated to the $R_{k}$ nodes will be
less noisy and, therefore, the skew and offset estimates made by a
node $ki$ will have less error. We would, thus, expect the variance
of the estimates to decrease with increasing $\bar{N}$. Notice that
our Type I network analysis does not explicitly utilize the circular
transmission region with radius $R$.

\subsection{Simulation Results} \label{sec:typeIsimulation}

\begin{figure}[!ht]
\centerline{\psfig{file=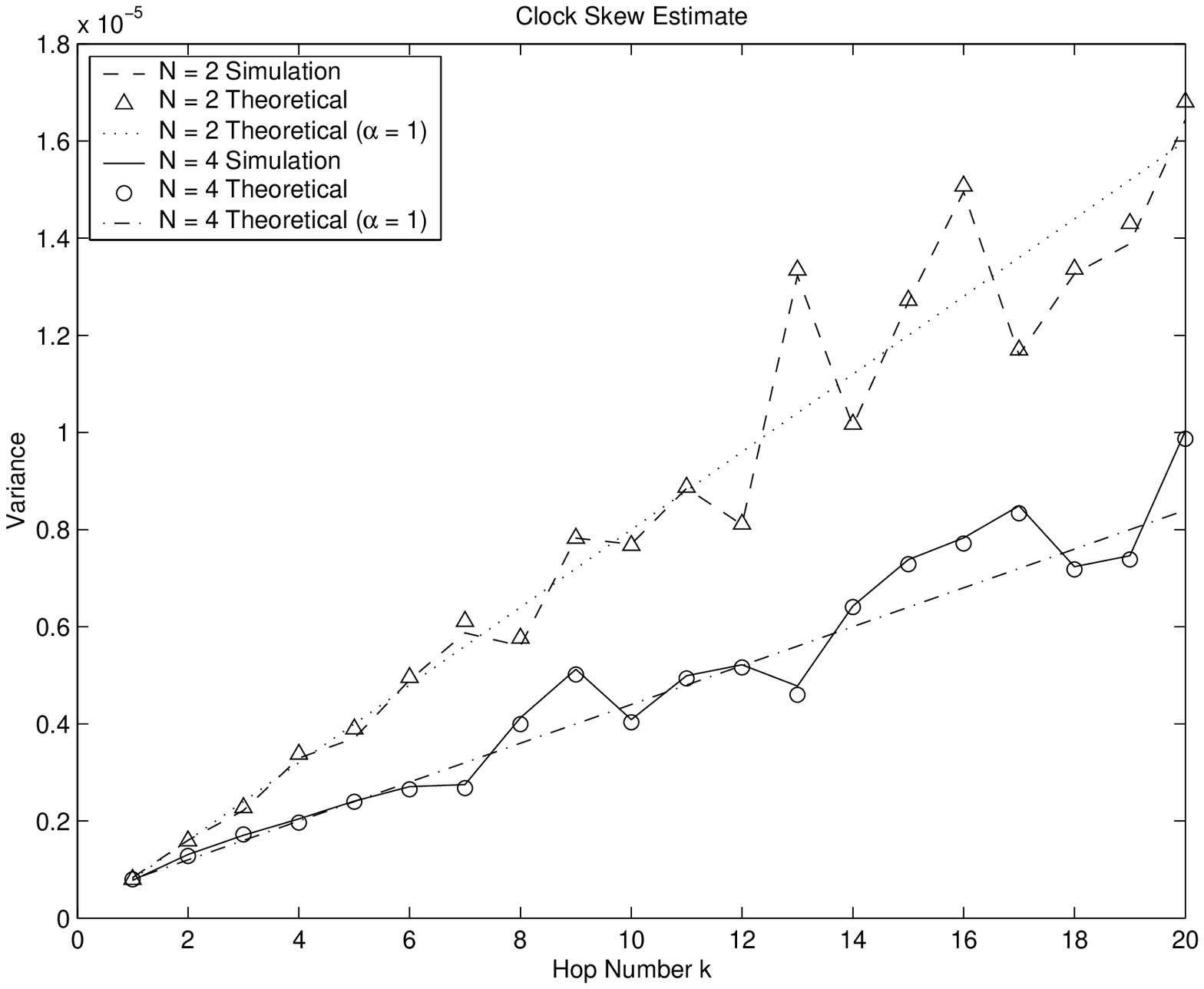,width=8.25cm}}
\centerline{\psfig{file=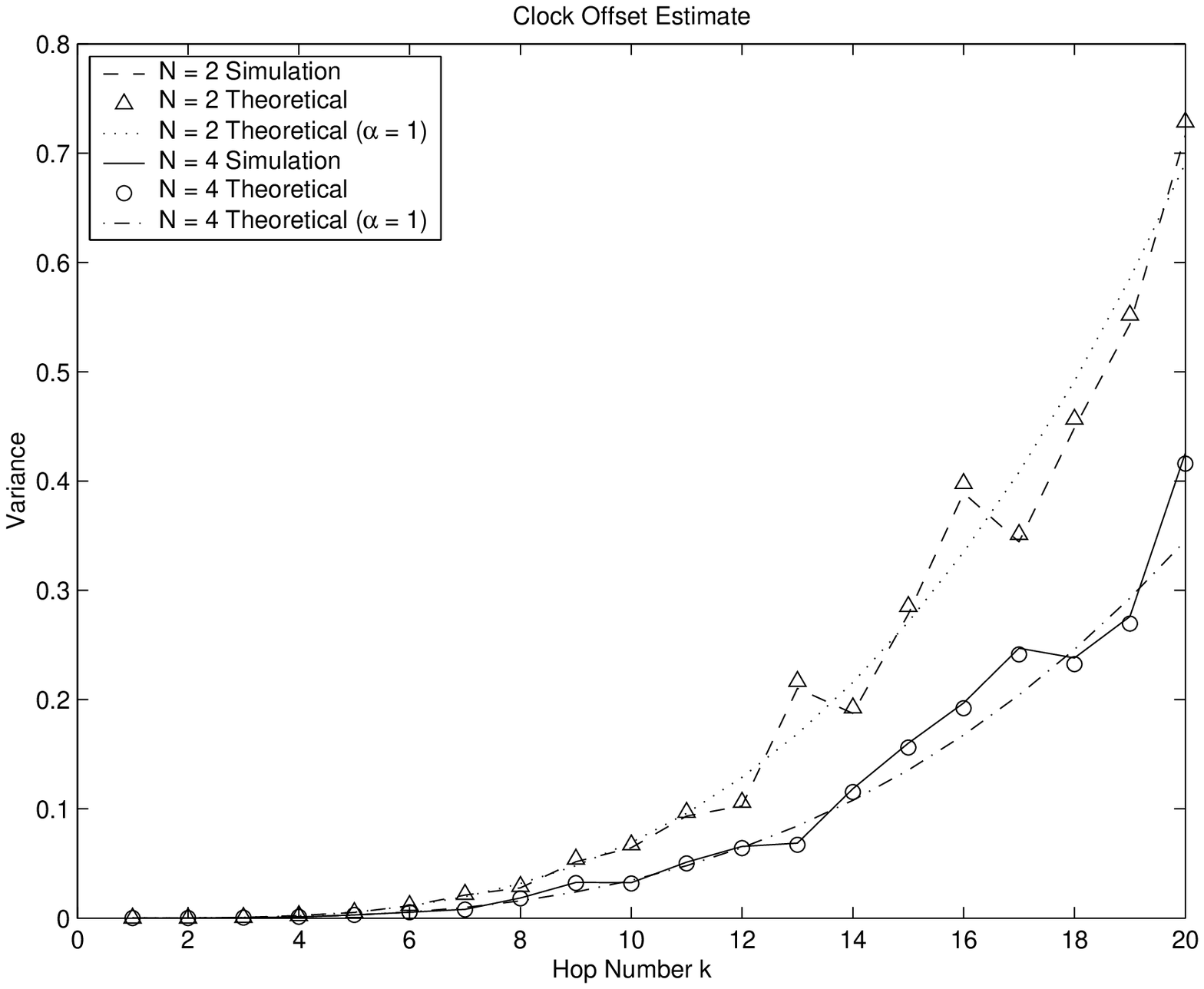,width=8.25cm}}
\caption{\small $\textrm{Var}(\hat{\alpha}_{k1})$ is plotted in the
top figure and $\textrm{Var}(\hat{\Delta}_{k1})$ is plotted in the
bottom figure as a function of $k$. }
\label{fig:typeI_alphaIncluded}
\end{figure}

In Fig.~\ref{fig:typeI_alphaIncluded} we illustrate the MATLAB
simulation results for two $20$ hop networks, one with $\bar{N}=2$
and the other with $\bar{N}=4$. The following parameters were used:
\begin{eqnarray*}
R = 1 \qquad d = 5 \qquad m = 4 \qquad \sigma = 0.01
\end{eqnarray*}
For each network, a set of $N=20\bar{N}+1$ nodes were first placed
in a Type I network deployment.  Each node's skew parameter was then
generated using $\alpha_{i} = |X_i|$ for $X_i\sim{\mathcal
N}(1,0.005)$, independently for each node $i$.  Node $1$ was assumed
to have $\alpha_1=1$.  The cooperative time synchronization protocol
was then run $5000$ times using the deployed network.  At each hop,
the $5000$ skew and offset estimates of one chosen node were used to
generate the simulated skew and offset estimate variance curves
shown in Fig.~\ref{fig:typeI_alphaIncluded}.  The theoretical
variance value of the chosen node at each hop was computed using the
recursive expression found in (\ref{eq:sigmak}).

In Fig.~\ref{fig:typeI_alphaIncluded}, we first clearly see that the
simulated skew and offset variance values nicely match the predicted
theoretical variance values.  As well, the expected decrease in skew
and offset variance as $\bar{N}$ increases from $2$ to $4$ is
immediately noticeable. In fact, in both the skew variance and
offset variance curves, we have an approximate halving of the
variance values as we double $\bar{N}$ from $2$ to $4$.  Also
expected, is that the variance values at each hop depend on the
particular values of $\alpha_{i}$, $i=1,\ldots,N$. This dependence
on the $\alpha_{i}$ values result in the jagged skew and offset
variance curves seen in Fig.~\ref{fig:typeI_alphaIncluded}. The
$\bar{N}=2$ network had $\alpha_{i}$ values ranging from $0.9073$ to
$1.1342$, while the $\bar{N}=4$ network had skew values ranging from
$0.8339$ to $1.1669$.

The problem with having the variance curves depend on the actual
skew values is that the exact performance of cooperative time
synchronization is dependent on the network realization. However, we
find that for $\alpha_{i}$ values that are close to and centered
around $1$, the variance curves follow the trend established by the
theoretical variance curves for $\alpha_{i}=1$, all $i$. This can be
seen in Fig.~\ref{fig:typeI_alphaIncluded} where we have also
plotted the theoretical curves using $\alpha_{i}=1$ for all $i$ for
$\bar{N}=2$ and $\bar{N}=4$. As a result, the situation where
$\alpha_{i}=1$, all $i$, can be used to study the the performance
improvement of cooperative time synchronization without dealing
specifically with the skew values of individual nodes.

Therefore, to get a better understanding of how cooperative time
synchronization improves synchronization performance, let us
simplify the recursive expression in (\ref{eq:sigmak}) for the
special case where $\alpha_{i}=1$ for all $i$ and find a
non-recursive expression for skew and offset variance.  The first
thing to note is that under the assumption of $\alpha_{i}=1$ for all
$i$, $\mathbf{A}_{k}=\mathbf{A}$ and $\Sigma_{m_{k}} = \Sigma_{m}$
are no longer dependent on $k$. Therefore, writing out the recursive
expression for $\bar{\Sigma}_{k}$ (\ref{eq:sigmak}), we have
\begin{equation} \label{eq:noalphaExpression}
\bar{\Sigma}_{k} =
\sum_{i=0}^{k-2}\mathbf{A}^{i}\Sigma_{m}(\mathbf{A}^{T})^{i} +
\mathbf{A}^{k-1}\bar{\Sigma}_{1}(\mathbf{A}^{T})^{k-1}.
\end{equation}
Using (\ref{eq:noalphaExpression}), Corollary~\ref{corollary:main}
gives us the non-recursive expression for skew and offset variance.

\begin{corollary} \label{corollary:main}
For a basic cooperative network with $\alpha_{i}=1$, all $i$,
$\hat{\alpha}_{ki}$ and $\hat{\Delta}_{ki}$ have the following mean
and variance:
\begin{equation}
E(\hat{\alpha}_{ki}) = 1 \nonumber
\end{equation}
\begin{equation}
E(\hat{\Delta}_{ki}) = -\bar{\Delta}_{ki} \nonumber
\end{equation}
\begin{equation} \label{eq:alphahatki_alphaEq1}
\textrm{Var}(\hat{\alpha}_{ki}) =
\frac{12\sigma^2}{d^2(m-1)m(m+1)}\left(
1+\frac{2(k-1)}{\bar{N}}\right)
\end{equation}
\begin{eqnarray} \label{eq:Deltahatki_alphaEq1}
\lefteqn{\textrm{Var}(\hat{\Delta}_{ki}) =
\frac{2\sigma^2(2m-1)}{m(m+1)} +
\frac{\sigma^2}{\bar{N}}\bigg[\frac{4(k-1)(2m-1)}{m(m+1)} }
\nonumber
\\
&&+ (k-1)^2\bigg(-\frac{12}{(m+1)}+\frac{12m}{(m-1)(m+1)}\bigg)
\nonumber \\
&&+ \frac{1}{3}(k-2)(k-1)(2k-3)\frac{12m}{(m-1)(m+1)}\bigg]
\end{eqnarray}
where $k$ is a positive integer. \qquad $\bigtriangleup$
\end{corollary}

The proof of Corollary~\ref{corollary:main} is omitted since it is a
direct simplification of (\ref{eq:noalphaExpression}). Note that the
skew and offset variance expressions are only a function of $k$ and
not $i$. The theoretical skew and offset variance of the $i$th node
at the $k$th hop (node $ki$) can be found in elements ($2(i-1)+2$,
$2(i-1)+2$) and ($2(i-1)+1$, $2(i-1)+1$), respectively, of
$\bar{\Sigma}_{k}$ in (\ref{eq:noalphaExpression}). However, the
skew variance values in elements ($2(i-1)+2$, $2(i-1)+2$),
$i=1,\ldots,\bar{N}$, are all equal and the offset variance values
in elements ($2(i-1)+1$, $2(i-1)+1$), $i=1,\ldots,\bar{N}$, are also
equal when we assume that $\alpha_{i}=1$ for all $i$.  As a result,
we can consider the skew and offset variance at a hop $k$ without
specifying a particular node.  Notice also that, besides the sign
change in the mean of the offset estimate, the skew and offset
estimates are unbiased estimates of the clock parameters of node
$ki$.

Looking at the skew and offset variance curves in
(\ref{eq:alphahatki_alphaEq1}) and (\ref{eq:Deltahatki_alphaEq1}),
respectively, we see that the variance growth decreases like
$1/\bar{N}$.  This $1/\bar{N}$ factor in both
(\ref{eq:alphahatki_alphaEq1}) and (\ref{eq:Deltahatki_alphaEq1}) is
expected since every node takes the sample mean of $\bar{N}$ pulses
to be an observation.  The variance of the observation decreases
like $1/\bar{N}$ because it is a sample mean and, thus, it is not
surprising that the skew and offset variance values also
approximately decrease like $1/\bar{N}$.

\section{Type II: General Networks} \label{sec:typeII}

\subsection{Network Setup}

Nodes will not generally be clustered together as in a basic
cooperative network, but be deployed in a more random manner. As a
result, to study general network deployments, we will consider a
Type II situation where nodes are uniformly deployed with density
$\rho$ over a circular region of radius $LR$ with node $1$ at the
center. In such a setup, at any hop $k$, $k\geq 2$, a node $ki$ in
the $R_{k}$ nodes will see \emph{at least} $\bar{N}$ nodes from the
$R_{k-1}$ set of nodes. However, the exact number of observed nodes
will depend on node $ki$'s location in the region occupied by the
$R_{k}$ nodes.

\begin{figure}[!ht]
\centerline{\psfig{file=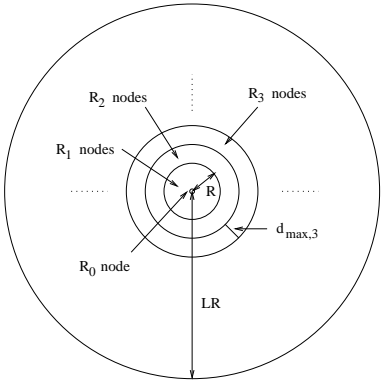,width=5cm}} \caption{\small A
Type II network deployment. Nodes are deployed with uniform density
$\rho$ and node $1$ is at the center of the network.}
\label{fig:type2setup}
\end{figure}

An illustration of a Type II deployment is shown in
Fig.~\ref{fig:type2setup}.  We note that the $R_{0}$ node (node $1$)
is placed at the center of the disk and the $R_{1}$ nodes occupy a
circular region of radius $R$.  However, the region occupied by the
$R_{k}$ nodes for $k\geq 2$ is a ring centered around node $1$ with
a ring thickness of $d_{max,k}$.  For increasing $k$, the distance
from node $1$ to the inner circular boundary of the region occupied
by the $R_{k}$ set of nodes increases.

\subsection{Analysis} \label{sec:typeII-analysis}

To study a Type II network, we could carry out an analysis similar
to the one we did for the Type I basic cooperative network. Assuming
we know the location of all nodes for a given network deployment
over the circular region of radius $LR$, we would be able to
determine the neighbors of each node and then readily extend the
Type I analysis to this Type II network.  The primary change that
would occur in the analysis is the determination of the affine
transform
\begin{displaymath}
\hat{\bar{\theta}}_{k} \mapsto
\bar{\theta}_{k+1}=\mathbf{A}_{k+1}\hat{\bar{\theta}}_{k}+\mathbf{B}_{k+1}.
\end{displaymath}
However, there are two issues that arise in determining the
transform matrix $\mathbf{A}_{k+1}$ and vector $\mathbf{B}_{k+1}$.

First, since $R_{k}$ and $R_{k+1}$ will most likely have different
numbers of nodes, we immediately see that $\mathbf{A}_{k+1}$ will be
a $2|R_{k+1}|\times 2|R_{k}|$ matrix and $\mathbf{B}_{k+1}$ will be
a $2|R_{k+1}|\times 1$ vector, where $|R_{k}|$ is the cardinality of
set $R_{k}$.  This means that the length of vector
$\hat{\bar{\theta}}_{k}$ will change with every hop.

Second, for any node $(k+1)i$ in $R_{k+1}$, the set of cooperating
nodes in $R_{k}$ will be different.  Thus, $\mathbf{A}_{k}$ will
also reflect this difference.  Therefore, every time we move from
hop $k$ to $k+1$, the correlation structure of
$\hat{\bar{\theta}}_{k}$ will change.

Together, these two points suggest that even though it is possible
to carry out the full analysis, the complexity would make the
resulting expressions depend on the particular network realization
and not provide significant insight into the problem.  In fact, it
would be nearly impossible to visualize the result without carrying
out a numerical evaluation.  Since our goal is to comprehend the
impact of spatial averaging on general networks, we choose to
proceed directly with simulations and compare the results with our
analytical expressions for Type I networks.

In the following analysis, we develop a basic understanding of what
we would expect to see in the simulation results that are presented
in Section~\ref{sec:typeIIIsimulation}. We assume that the number of
nodes in any given area of the Type II network is proportional to
the area. The reason is that for uniformly deployed nodes with
density $\rho$, the average number of nodes in an area ${\mathcal
A}$ is ${\mathcal A}\rho$.  Note that even though the analysis and
simulation results for Type II networks use the assumption of a
circular transmission range of $R$, the simulation results in
Section~\ref{sec:typeIIIsimulation} still provide valid insight when
realistic transmission regions~\cite{GanesanKWCEW:02,ZhaoG:03} are
assumed since the figures illustrate synchronization error as a
function of \emph{hop} number.  Therefore, regardless of the shape
of the transmission region, a node at hop $k$ will have received the
appropriate synchronization information and, thus, our simulation
results reflect its synchronization performance.

\subsubsection{Estimation of $\bar{L}$} \label{sec:estimatebarL}

\begin{figure}[!ht]
\centerline{\psfig{file=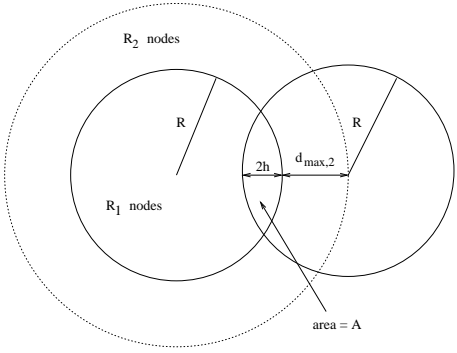,width=6cm}} \caption{\small An
illustration of $d_{max,2}$.} \label{fig:type2dmax}
\end{figure}

Our first consideration is to estimate the number of hops,
$\bar{L}$, required to communicate timing information from node $1$
to the edge of the network a distance $LR$ away. In order to do
this, we need a way to quantify $d_{max,k}$. In
Fig.~\ref{fig:type2dmax}, we illustrate $d_{max,2}$ and see that
$d_{max,2}$ is determined by having the intersection of the two
radius $R$ circles contain an average of $\bar{N}$ nodes. This is
because if we increase $d_{max,2}$, then nodes at this increased
distance will not see $\bar{N}$ nodes on average and, thus, not be
considered an $R_{2}$ node.  However, $d_{max,k}>d_{max,2}$, for
$k>2$, because the ring occupied by the $R_{k}$ nodes increases in
size for increasing $k$.  As a result, we choose to be conservative
and let $d_{max}\stackrel{\Delta}{=}d_{max,2}$ approximate
$d_{max,k}$ for all $k$. This means that our estimate of $\bar{L}$
using $d_{max}$ will be greater than or equal to the number of hops
required to reach a distance of $LR$ when the differences in
$d_{max,k}$ are considered.

Let $A$ be the area of the intersection of the two radius $R$
circles in Fig.~\ref{fig:type2dmax} and we have
from~\cite{Weisstein:03} that
\begin{equation} \label{eq:circsegarea}
A =
2\left(R^2\cos^{-1}\left(\frac{R-h}{R}\right)-(R-h)\sqrt{2Rh-h^2}\right).
\end{equation}
Since $A$ contains $\bar{N}$ nodes, we have that
\begin{equation} \label{eq:Aexpression}
A = \bar{N}/\rho.
\end{equation}
From (\ref{eq:circsegarea}) and (\ref{eq:Aexpression}) we can
numerically determine $h$ thus giving us
\begin{equation} \label{eq:dmax2}
d_{max} = R-2h.
\end{equation}
As a result, we need $\bar{L}$ to satisfy
\begin{displaymath}
R+(\bar{L}-1)d_{max}\geq LR
\end{displaymath}
which means that
\begin{equation} \label{eq:barL2}
\bar{L} = \left\lceil\frac{R(L-1)}{R-2h}+1 \right\rceil.
\end{equation}

\subsubsection{Comparison to Type I Networks} \label{sec:uplowbounds}

We will compare the Type II network simulation results to the Type I
analytical results.  This comparison will allow us to carry over the
intuition regarding spatial averaging that we have developed for the
basic cooperative network.  However, Type I and Type II networks
differ primarily in that Type I networks assume that all nodes will
observe $\bar{N}$ neighbors from the previous hop while any node in
a Type II network will only see \emph{at least} $\bar{N}$ nodes.
Thus, if we want to compare Type I and Type II plots, we need to
establish some meaningful choices of the number of cooperating nodes
for use with expressions (\ref{eq:alphahatki_alphaEq1}) and
(\ref{eq:Deltahatki_alphaEq1}).

\begin{figure}[!ht]
\centerline{\psfig{file=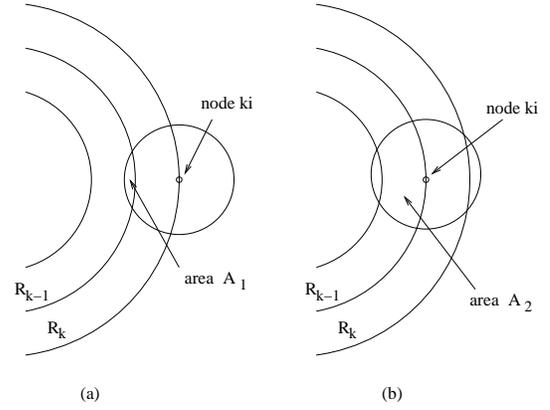,width=7cm}} \caption{\small
(a) Node $ki$ at the outer circular boundary of the $R_{k}$ set of
nodes.  (b) Node $ki$ at the inner circular boundary of the $R_{k}$
set.} \label{fig:type2Nbark}
\end{figure}

\begin{figure}[!ht]
\centerline{\psfig{file=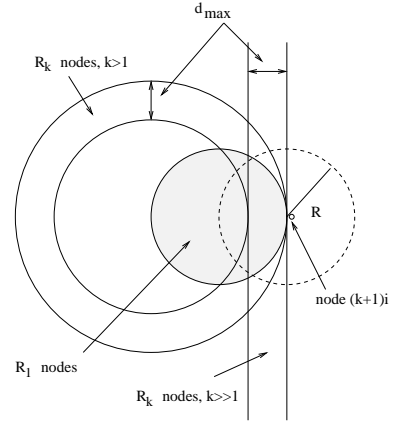,width=5cm}} \caption{The
regions occupied by the $R_{k}$ nodes for $k=1$, $k>1$, and $k>>1$
overlayed on top of each other. The region of nodes seen by a node
at the inner circular boundary of $R_{k+1}$ changes with $k$. \small
} \label{fig:type2Nmax}
\end{figure}

Looking at (a) of Fig.~\ref{fig:type2Nbark}, we see that if a node
$ki$ in the region occupied by the $R_{k}$ nodes is at the circular
boundary farthest from node $1$ (outer circular boundary), then it
will likely hear only $\bar{N}$ nodes from $R_{k-1}$. That is, there
are $\bar{N}=A_{1}\rho$ nodes in area $A_{1}$. Recall that $\bar{N}$
is the minimum number of $R_{k-1}$ nodes any node $ki$ will hear.
However, looking at (b) in Fig.~\ref{fig:type2Nbark}, a node $ki$ at
the circular boundary closest to node $1$ (inner circular boundary)
in the $R_{k}$ region will hear many more nodes. In fact, a node
$ki$ at the boundary between $R_{k-1}$ and $R_{k}$ will hear the
largest average number of nodes $\bar{N}_{max}(k)=A_{2}\rho$. Since
$\bar{N}$ and $\bar{N}_{max}(k)$ is the range of the number of
cooperating nodes seen by a node in $R_{k}$, it would make sense to
plot Type I expressions (\ref{eq:alphahatki_alphaEq1}) and
(\ref{eq:Deltahatki_alphaEq1}) using these two values. However,
$\bar{N}_{max}(k)$ varies with $k$.  In Fig.~\ref{fig:type2Nmax} we
illustrate the regions occupied by the $R_{k}$ nodes for $k=1$,
$k>1$, and $k>>1$ overlayed on top of each other and in each
situation, we see that the set of nodes in $R_{k}$ seen by a node at
the boundary between the $R_{k}$ nodes and the $R_{k+1}$ nodes is
different for changing values of $k$. However, it is clear that the
area of intersection always falls inside a semicircle of radius $R$.
As a result, we will approximate $\bar{N}_{max}=\max_{k:k\geq
2}\bar{N}_{max}(k)$, by upper bounding the maximum area of
intersection with the area of the semicircle. This means that
\begin{equation} \label{eq:nbarmax2}
\bar{N}_{max} \approx \rho\frac{\pi R^{2}}{2}.
\end{equation}
Thus, in comparing Type II and Type I results, we will use $\bar{N}$
and $\bar{N}_{max}$ in (\ref{eq:nbarmax2}) with both
(\ref{eq:alphahatki_alphaEq1}) and (\ref{eq:Deltahatki_alphaEq1})

Using $\bar{N}$ with (\ref{eq:alphahatki_alphaEq1}) and
(\ref{eq:Deltahatki_alphaEq1}) will provide a curve that tends to be
higher than the Type II simulated curves for two main reasons.
First, since $\bar{N}$ is the minimum number of nodes in $R_{k-1}$
that a node $ki$ in $R_{k}$ will hear and we know that a larger
number of cooperating nodes will result in decreased estimation
variance, the variance values computed using $\bar{N}$ will tend to
be higher. Second, even if a node $ki$ in $R_{k}$ hears $\bar{N}$
nodes from $R_{k-1}$, each of those $\bar{N}$ nodes did not
necessarily only hear $\bar{N}$ nodes from $R_{k-2}$. Thus, the skew
and offset estimates made by each of those $\bar{N}$ nodes in
$R_{k-1}$ whose transmissions are being heard by node $ki$ may have
a variance that is less than predicted by
(\ref{eq:alphahatki_alphaEq1}) and (\ref{eq:Deltahatki_alphaEq1})
using $\bar{N}$. The improved skew and offset estimates made by the
nodes in $R_{k-1}$ will thus lead to a lower estimation variance for
node $ki$ even though node $ki$ hears only $\bar{N}$ from $R_{k-1}$.

Using $\bar{N}_{max}$ with (\ref{eq:alphahatki_alphaEq1}) and
(\ref{eq:Deltahatki_alphaEq1}) will provide a curve that tends to be
lower than the Type II simulated curves for two similar reasons.
First, since $\bar{N}_{max}$ is the average number of nodes heard by
a node at the inner circular boundary of $R_{k}$, $k\geq2$, and all
other nodes in $R_{k}$ will on average hear fewer nodes, a Type I
curve using $\bar{N}_{max}$ will tend to yield lower values. Second,
not all nodes in $R_{k-1}$ make their estimates using a signal
cooperatively generated by $\bar{N}_{max}$ nodes.  In fact, most
nodes in $R_{k-1}$ observe fewer than $\bar{N}_{max}$ nodes. As a
result, the lower quality estimates made by some of the $R_{k-1}$
nodes will cause the estimation variance of the $R_{k}$ nodes that
hear $\bar{N}_{max}$ from $R_{k-1}$ to be greater than predicted by
(\ref{eq:alphahatki_alphaEq1}) and (\ref{eq:Deltahatki_alphaEq1})
using $\bar{N}_{max}$.

\subsubsection{Synchronization Performance and Node Density}
\label{sec:perfandnodedensity}

The third issue we want to address in analyzing a Type II network
deployment is how to decrease synchronization error when we know
from Section \ref{sec:estimatebarL} that the number of hops
$\bar{L}$ required to communicate timing information from node $1$
to the edge of the network a distance $LR$ away is determined by
$\bar{N}$. Given a fixed $R$, we can start with some $\bar{N}$ and
$\rho$. Using (\ref{eq:circsegarea}), (\ref{eq:Aexpression}), and
(\ref{eq:dmax2}), we can determine the value of $d_{max}$ and,
hence, from (\ref{eq:barL2}) the number of hops $\bar{L}$ required
to send timing information from node $1$ to the edge of the network.
In order to decrease synchronization error at a distance $LR$ from
node $1$, we need to increase $\bar{N}$. However, only increasing
$\bar{N}$ will decrease $d_{max}$ and increase $\bar{L}$. Therefore,
we need to increase both $\bar{N}$ and $\rho$. From
(\ref{eq:circsegarea}) and (\ref{eq:Aexpression}), we see that if
$\bar{N}/\rho$ is kept constant, then $h$ will be constant.  If $h$
is constant, then so is $d_{max}$. As a result, by increasing node
density, we can increase the minimum number of cooperating nodes
$\bar{N}$ and therefore decrease synchronization error.

\subsection{Simulation Results} \label{sec:typeIIIsimulation}

In the following simulation results, we have assumed that all nodes
in the network have no clock skew, i.e. $\alpha_{i}=1$ for all $i$.
From Section~\ref{sec:typeIsimulation} we know that general
$\alpha_{i}$ values result in variance curves that follow the trends
established by curves generated using $\alpha_{i}=1$. As a result,
using $\alpha_{i}=1$ for all $i$ allows us to study the benefits of
spatial averaging without considering effects that are dependent on
the particular network realization.

\subsubsection{Comparison to Type I Results}
\label{sec:typeIIIsimulation-bounds}

\begin{figure}[!ht]
\centerline{\psfig{file=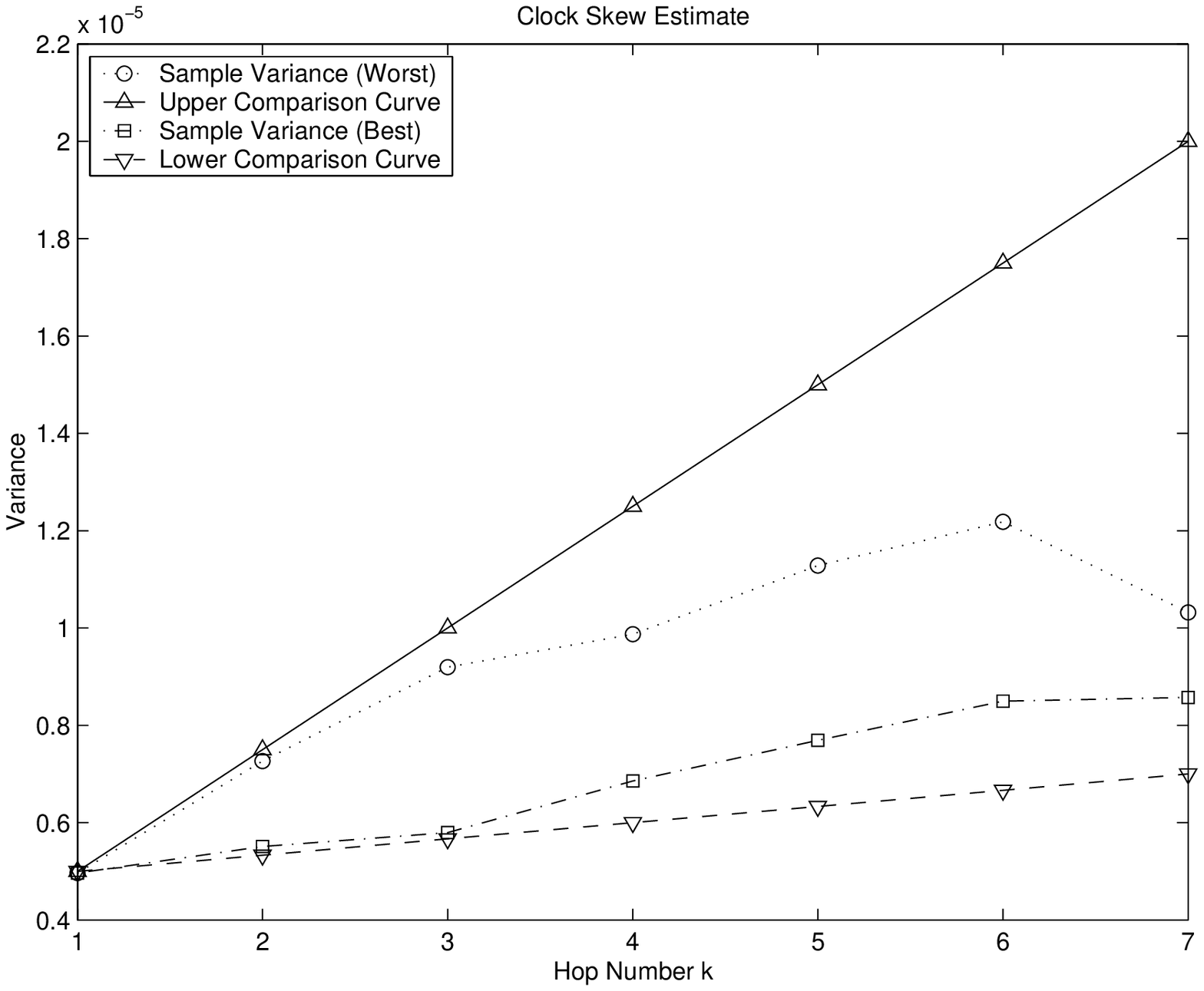,width=8.25cm}}
\centerline{\psfig{file=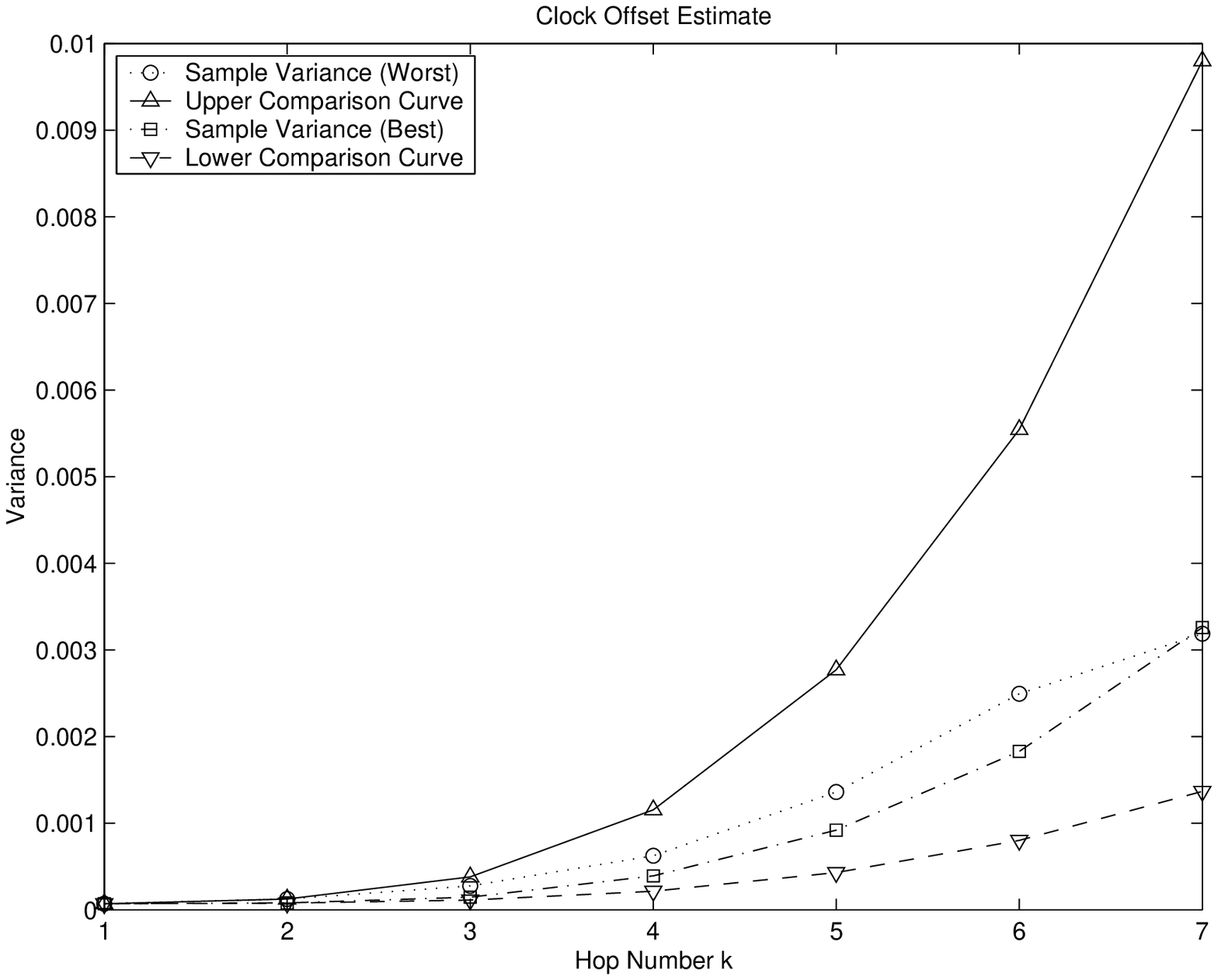,width=8.25cm}}
\caption{\small Simulation $1$. Top: Sample variance for the skew
estimate of a Type II network along with Type I comparison curves.
Bottom: Sample variance of the offset estimate along with Type I
comparison curves.} \label{fig:type2bound}
\end{figure}

To being the study of cooperative time synchronization in general
networks, we deploy a network for Simulation $1$ with the parameters
in Table~\ref{tab:simulation1}.
\begin{table}[!ht]
\caption{\rm Simulation 1 Parameters}
\begin{center}
\begin{tabular}{|c|c|}
\hline
$\rho$ & 19.10 \\
$\bar{N}$ & 4 \\
$R$ & 1 \\
$L$ & 5 \\
$d$ & 2 \\
$m$ & 4 \\
$\sigma$ & 0.01 \\
Number of Runs & 5000 \\
\hline
\end{tabular}
\end{center}
\label{tab:simulation1}
\end{table}
The simulation results are displayed in Fig.~\ref{fig:type2bound}.
In each run, a new network of nodes was uniformly deployed over a
circular area of radius $LR=5$ and the MATLAB simulator implemented
the cooperative time synchronization protocol. Besides plotting the
Type I comparison curves described in Section~\ref{sec:uplowbounds},
we also plot the sample variance of the best performing node and the
worse performing node. In each run, the node in $R_{k}$ that sees
the fewest number of nodes from $R_{k-1}$ is considered the worse
performing node while the node in $R_{k}$ that sees the largest
number of nodes from $R_{k-1}$ is the best performing node. For the
$l$th run, the fewest number of nodes seen by a node in $R_{k}$ is
denoted $X_{min}^{(l)}(k)$ while the largest number of nodes seen by
a node in $R_{k}$ is denoted $X_{max}^{(l)}(k)$.  The skew and
offset estimate of the best and worst performing node at each hop is
recorded and the sample variance over the $5000$ runs is plotted.

The top figure in Fig.~\ref{fig:type2bound} illustrates the sample
skew variance curves of the worst and best synchronized node along
with the Type I curves for comparison. The bottom figure in
Fig.~\ref{fig:type2bound} illustrates the clock offset estimate
sample variance. Note that using equation (\ref{eq:barL2}) and the
parameters in Table~\ref{tab:simulation1}, we find that $\bar{L}=7$.
From the simulations, we also see that $7$ hops are required to
traverse the network. In fact, only $7.32\%$ of the networks
required more than $7$ hops to reach all nodes in the network.

As predicted in Section~\ref{sec:uplowbounds}, we clearly see in
Fig.~\ref{fig:type2bound} that the worst case variance and the best
case variance are sandwiched between the Type I comparison curves.
Also, as expected, the skew and offset variances do not closely
follow the upper and lower Type I comparison curves.  The worst case
skew and offset variance follow the upper comparison cruve for the
first $2$ hops and then begin do deviate from the curve.  As
mentioned in Section~\ref{sec:uplowbounds}, this is because the
nodes contributing to the worst performing node may have received
signals from more than $\bar{N}$ nodes. Similarly, the best case
skew and offset variance follow the lower comparison curve for the
first $2$ hops before deviating. This is because many of the nodes
contributing signals to the best performing node made their
estimates using a signal cooperatively generated by less than
$\bar{N}_{max}$ nodes. Also of interest is the steep decrease in the
worst case skew and offset variance at hop $k=7$.  This is due to
the fact that on average, the distance from the outer circular
boundary of the $R_{6}$ region to the network boundary is much less
than $d_{max}$. As a result, the $R_{7}$ region is smaller and
$X_{min}(7)$ will be larger than $\bar{N}$.
Table~\ref{tab:type2_nbarvalues1} shows the
$X_{min}(k)=\frac{1}{5000}\sum_{l=1}^{5000}X_{min}^{(l)}(k)$ and
$X_{max}(k)=\frac{1}{5000}\sum_{l=1}^{5000}X_{max}^{(l)}(k)$ values
and we see that $X_{min}(6)=\bar{N}=4$, but $X_{min}(7)$ is nearly
twice $X_{min}(6)$.

\begin{table}[!ht]
\caption{\rm $X_{min}(k)$ and $X_{max}(k)$ for
Fig.~\ref{fig:type2bound}}
\begin{center}
\begin{tabular}{|c|c|c|}
\hline
$k$& $X_{min}(k)$ & $X_{max}(k)$\\
\hline
1&1&1\\
2&4.00&27.56\\
3&4.00&29.36\\
4&4.00&31.86\\
5&4.00&33.50\\
6&4.00&34.60\\
7&7.77&35.32\\
\hline
\end{tabular}
\end{center}
\label{tab:type2_nbarvalues1}
\end{table}

We also note that $X_{max}(k)$ increases from $27.56$ for $k=2$ to
$35.32$ for $k=7$.  Using (\ref{eq:nbarmax2}), however, we find that
$\bar{N}_{max}=30$.  The reason $X_{max}(k)$ increases with each hop
and does not equal $\bar{N}_{max}$ is because $X_{max}(k)$ is a
different statistic.  $\bar{N}_{max}$ approximates the average
number of $R_{k-1}$ nodes seen by a node $ki$ at the inner circular
boundary of $R_{k}$. However, $X_{max}^{(l)}(k)$ is \emph{the}
largest number of nodes seen by any node $ki$ in $R_{k}$ for the
$l$th network realization. Therefore, $X_{max}^{(l)}(k)$ is actually
an ordered statistic since it takes the largest number of nodes seen
by a node at hop $k$. $X_{max}(k)$ is thus the mean of the ordered
statistic. Therefore, we would not expect $\bar{N}_{max}$ and
$X_{max}(k)$ to be the same. Also, $X_{max}(k)$ increases with $k$
since as the circumference of the circular ring occupied by $R_{k}$
increases, there are more nodes at the boundary between $R_{k}$ and
$R_{k-1}$.  Since there are more nodes at the boundary, there are
also more opportunities to find the largest number of nodes seen by
a node $ki$.  Thus, the maximum number of nodes would tend to be
larger. Note that, not considering the effects at the network
boundary, $X_{min}(k)=\bar{N}$ because the definition of the
protocol specifies the minimum to be $\bar{N}$ and there is little
randomness in determining $X_{min}(k)$.

\subsubsection{Synchronization Performance and Node Density}
\label{sec:typeIIIsimulation-densityincrease}

\begin{figure}[!ht]
\centerline{\psfig{file=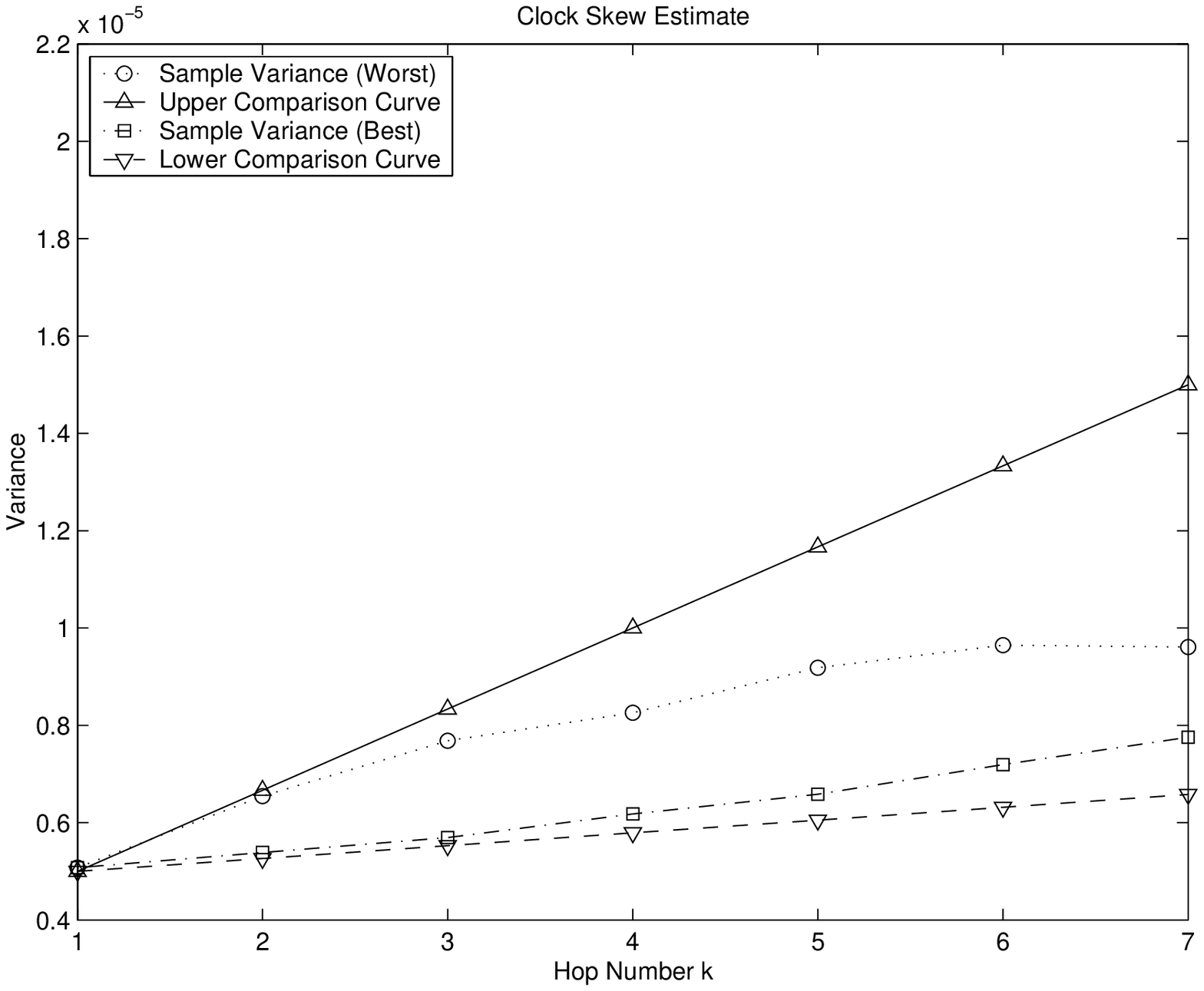,width=8.25cm}}
\centerline{\psfig{file=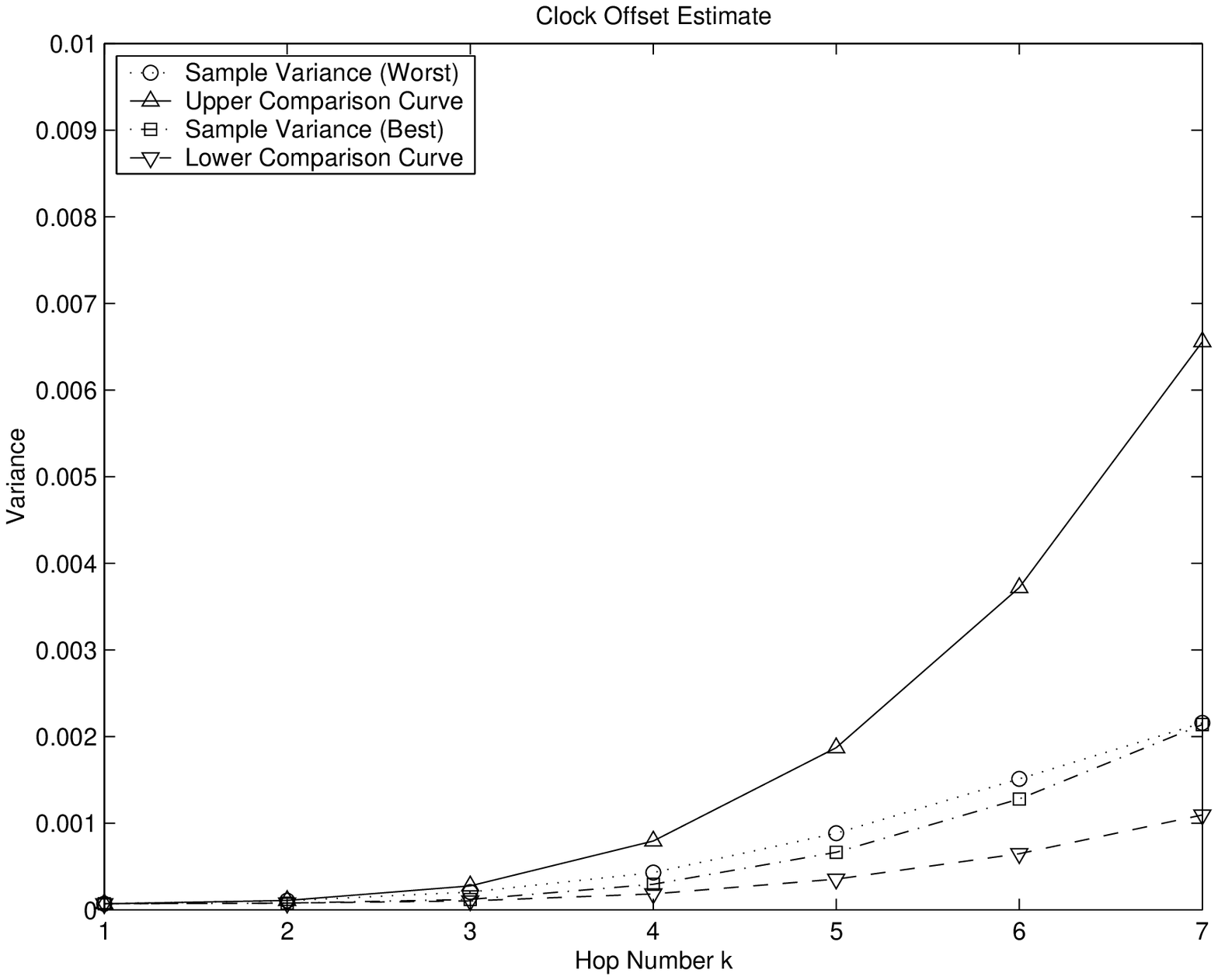,width=8.25cm}}
\caption{\small Simulation $1$b. The Type I comparison curves and
the Type II sample skew and offset variance curves are lower as
compared to Fig.~\ref{fig:type2bound} when $\bar{N}$ and $\rho$ are
increased. More cooperation yields improved synchronization
performance.} \label{fig:type2bound_lower}
\end{figure}

\begin{table}[!ht]
\caption{\rm Simulation 1b Parameters}
\begin{center}
\begin{tabular}{|c|c|}
\hline
$\rho$ & 23.87 \\
$\bar{N}$ & 6 \\
$R$ & 1 \\
$L$ & 5 \\
$d$ & 2 \\
$m$ & 4 \\
$\sigma$ & 0.01 \\
Number of Runs & 5000 \\
\hline
\end{tabular}
\end{center}
\label{tab:simulation1b}
\end{table}

Next, we want to improve synchronization performance by increasing
node density.  Starting with the parameters for Simulation $1$, we
increase the minimum number of cooperating nodes to $\bar{N}=6$
while keeping $\bar{N}/\rho=0.25$ constant.  Therefore, for
Simulation $1$b (Table~\ref{tab:simulation1b}), $\rho=23.87$ and we
plot the simulation results in Fig.~\ref{fig:type2bound_lower}.
Comparing Fig.~\ref{fig:type2bound} and
Fig.~\ref{fig:type2bound_lower}, it is clear that
Fig.~\ref{fig:type2bound_lower} yields improved skew and offset
variances, thus showing that increased node density and larger
$\bar{N}$ values indeed improve synchronization performance in Type
II networks. In Table~\ref{tab:type2_nbarvalues2} we show
$X_{min}(k)$ and $X_{max}(k)$ for Fig.~\ref{fig:type2bound_lower}.
Note that there is only a slight decrease in the worst case skew and
offset variance curves at hop $k=7$ since in this simulation, we
have that $X_{min}(7)=6.57$ is only slightly larger than
$X_{min}(6)=\bar{N}=6$.

\begin{table}[!ht]
\caption{\rm $X_{min}(k)$ and $X_{max}(k)$ for
Fig.~\ref{fig:type2bound_lower}}
\begin{center}
\begin{tabular}{|c|c|c|}
\hline
$k$& $X_{min}(k)$ & $X_{max}(k)$\\
\hline
1&1&1\\
2&6.00&34.01\\
3&6.00&34.64\\
4&6.00&37.64\\
5&6.00&39.50\\
6&6.00&40.80\\
7&6.57&41.70\\
\hline
\end{tabular}
\end{center}
\label{tab:type2_nbarvalues2}
\end{table}

\begin{figure}[!ht]
\centerline{\psfig{file=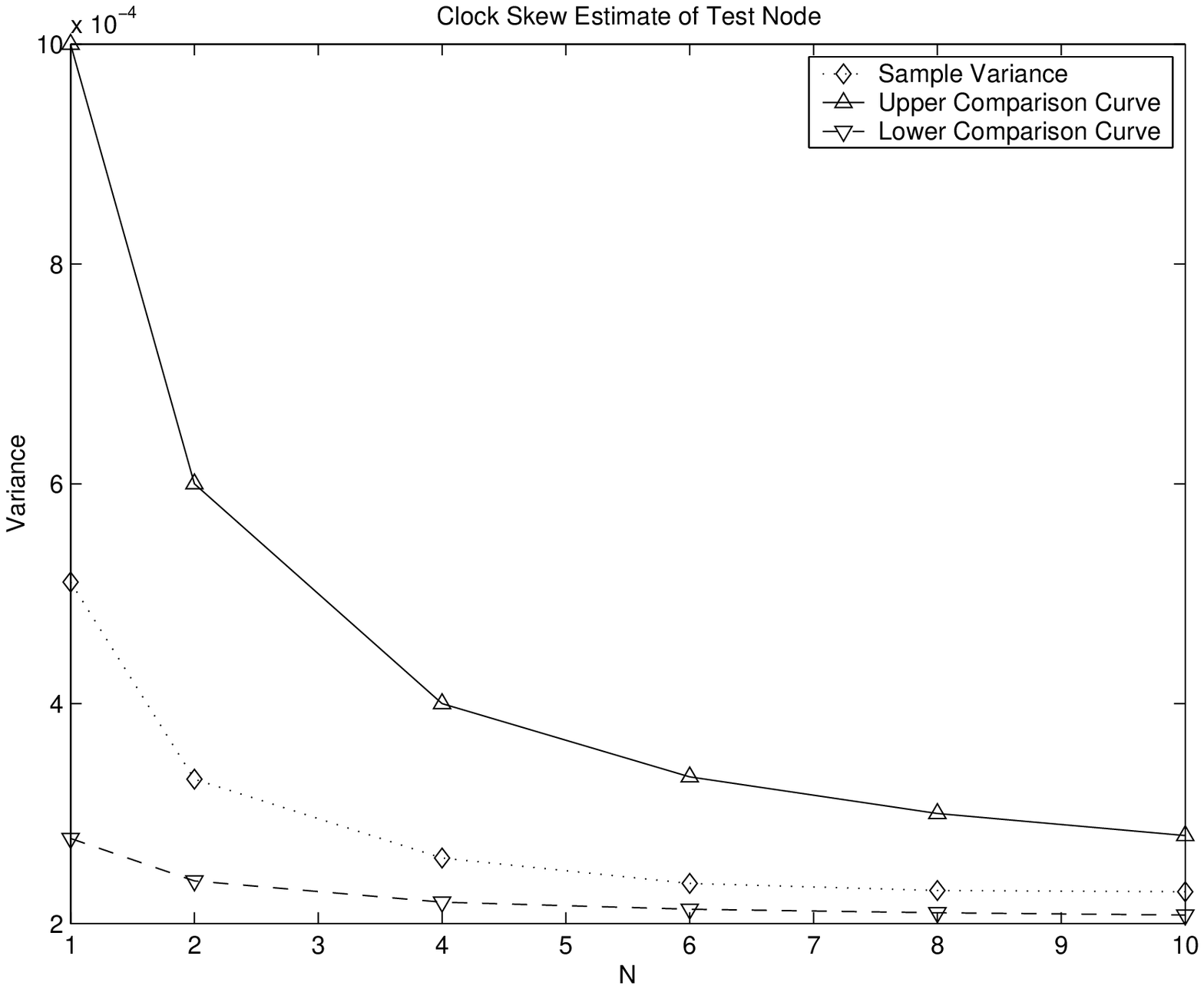,width=8.25cm}}
\centerline{\psfig{file=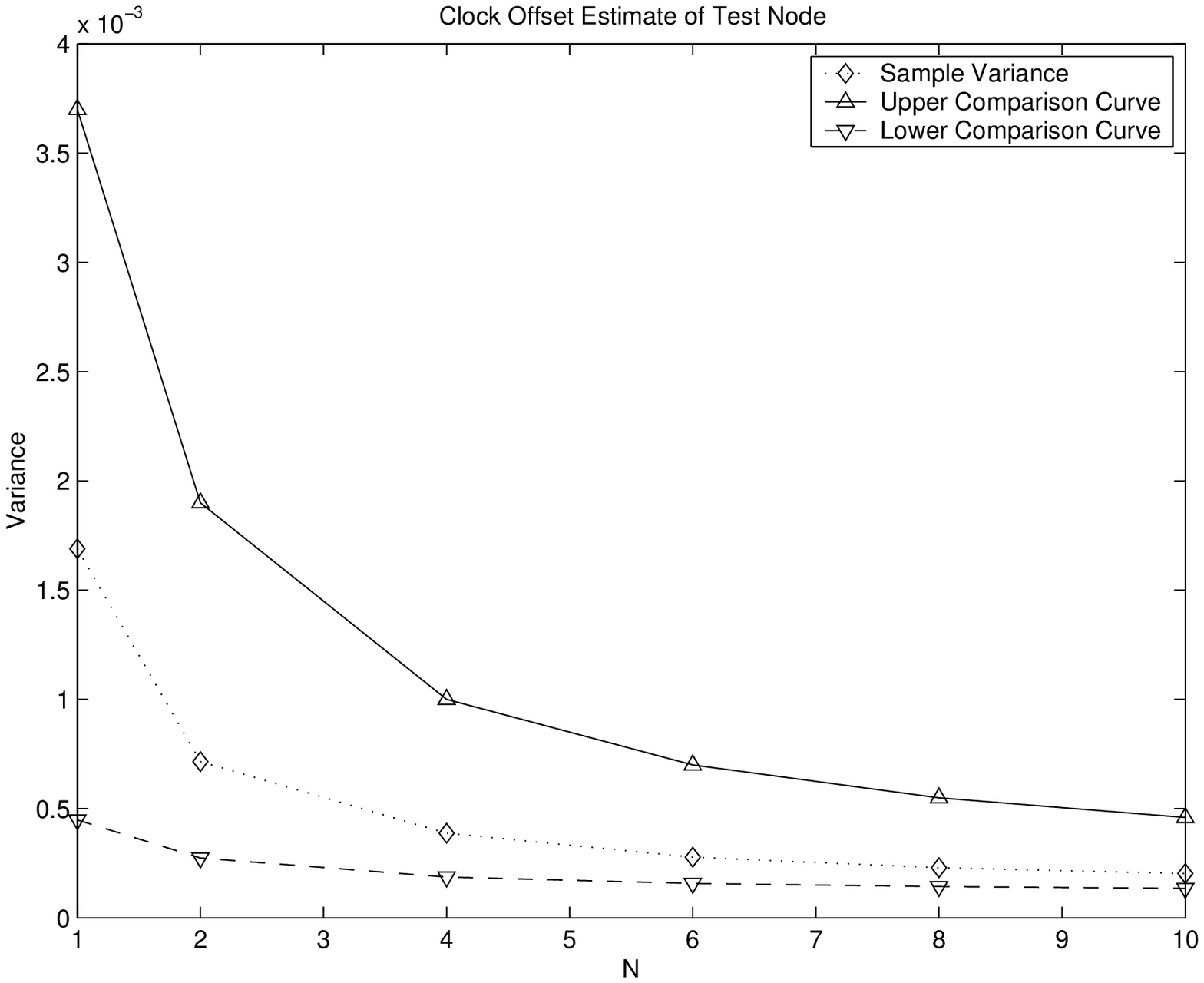,width=8.25cm}}
\caption{\small Simulation $2$.  Variance of the skew and offset
estimates of the test node fall between the Type I comparison curves
and decrease with increasing $\bar{N}$ and $\rho$.}
\label{fig:type2variancedecrease}
\end{figure}

Another very effective way to visualize how increasing $\rho$ and
$\bar{N}$ can decrease skew and offset variance is to choose one
\emph{test node} in the network and consider how its skew and offset
variance decreases as the network density and number of cooperating
nodes are increased. In Simulation 2 (Table~\ref{tab:simulation3}),
we placed a test node at distance $LR=2.2$ from node $1$ and
simulated its skew and offset variance as we increased $\rho$ and
$\bar{N}$.  $\bar{N}$ took on values ranging from $1$ to $10$ and we
adjusted $\rho$ accordingly to keep $\bar{N}/\rho=0.15$ fixed.  The
results are plotted in Fig.~\ref{fig:type2variancedecrease} and we
clearly see that as $\bar{N}$ increases along with $\rho$, the skew
and offset variance of this test node decreases.  Also, from
Section~\ref{sec:perfandnodedensity}, we know that since we keep
$\bar{N}/\rho$ constant, the number of hops required to reach the
test node stays the same as we increase $\bar{N}$.  Therefore, since
the test node is at $\bar{L}=3$ for every value of $\bar{N}$, we
have also plotted the upper and lower Type I comparison curves for
the skew and offset variance at hop $k=3$ to illustrate how the
comparison curves change in relation to the simulated variance
curves.  In Fig.~\ref{fig:type2variancedecrease}, the simulated skew
and offset variance curves of the test node fall between the Type I
upper and lower comparison curves.

\begin{table}[!ht]
\caption{\rm Simulation 2 Parameters}
\begin{center}
\begin{tabular}{|c|c|}
\hline
$\bar{N}/\rho$ & 0.15 \\
$\bar{N}$ & [1 2 4 6 8 10] \\
$R$ & 1 \\
$L$ & 2.2 \\
$d$ & 1 \\
$m$ & 2 \\
$\sigma$ & 0.01 \\
Number of Runs & 5000 \\
\hline
\end{tabular}
\end{center}
\label{tab:simulation3}
\end{table}

It is clear that by keeping the ratio $\bar{N}/\rho$ constant while
increasing $\bar{N}$ and $\rho$ allows us to reduce the
synchronization error at each hop while keeping the number of hops
required to synchronize the network, $\bar{L}$, constant.  The
variance of the skew and offset estimates is decreased by increasing
the minimum number of cooperating nodes.

Furthermore, from the simulations in this section, we find that the
upper and lower Type I comparison curves provide a good reference to
the performance of Type II networks.  We have established that the
best and worst case variance values for the Type II skew and offset
estimates fall between the upper and lower Type I comparison curves.
As the density of the network and $\bar{N}$ are both increased, the
comparison curves will shift downwards and become closer together.
Thus, we would expect the variance of the Type II network estimates
to change similarly with increasing $\bar{N}$ and $\rho$.

\section{Conclusion} \label{sec:conclusion}

In this paper we have proposed one technique that uses spatial
averaging in dense networks as a means to improving global time
synchronization. Spatial averaging is used to improve the timing
data points that are used to estimate clock skew and clock offset.
By decreasing the error in the timing data points, improved clock
skew and clock offset estimates can be made. Our analysis of the
technique in a basic cooperative network revealed that the error
variance in both the clock skew and clock offset estimates can be
significantly decreased as the number of cooperating nodes
increases.  Simulation results also show that synchronization over
large, multi-hop networks can be improved by increasing node
density.  Further analysis and a comparison between cooperative and
non-cooperative techniques can be found in~\cite{Hu:07}.

This scalable protocol provides an alternate way to combat the
scalability problem. It allows us to simply increase the number of
nodes in the network to obtain improved synchronization performance.
The new trade-off between network density and synchronization
performance provided by spatial averaging will allow for added
flexibility in designing future networks.

It is important to note that the concept of spatial averaging is
very general and our proposed cooperative technique is but one
manner in which to take advantage of it.  Our protocol shows that
techniques using spatial averaging can be designed.  Even though the
proposed protocol has certain limitations, such as requiring access
to the physical layer, it allows us to successfully illustrate the
performance improvement achievable using spatial averaging.  Future
work will focus on other approaches to spatial averaging. For
example, it would be desirable to develop a cooperative technique
using spatial averaging that achieves performance gains while
needing only access to the data link or network layer.

\appendix

{\bf Proof of Theorem~\ref{theorem:main}} \hspace{0.5cm} Node $1$
begins the synchronization processes by transmitting a sequence of
pulses at times $\tau_{0}+ld$, for $l=0,\ldots,m-1$. For simplicity,
assume that $\tau_{0}$ and $d$ are integer values. Note that since
node $1$ transmits these pulses in its own time scale $c_{1}$ (the
reference time), the pulses will occur at integer values of $t$.
Using the clock model in (\ref{eq:clock}), any node $1i$,
$i=1,\ldots,\bar{N}$, in the $R_{1}$ set of nodes will get a vector
of observations ${\mathbf Y}_{1i}$, where ${\mathbf Y}_{1i}[1] =
\alpha_{1i}(\tau_{0}-\bar{\Delta}_{1i})+\Psi_{1i,1}$ and the
$(l+1)$th element of ${\mathbf Y}_{1i}$ is ${\mathbf
Y}_{1i}[l+1]=\alpha_{1i}(\tau_{0}-\bar{\Delta}_{1i})+ld\alpha_{1i}+\Psi_{1i,l+1}$.
This can also be written as
\begin{equation} \label{eq:linearFormCoop}
{\mathbf Y}_{1i}={\mathbf H}\theta_{1i} + {\mathbf W}_{1i},
\end{equation}
where
\begin{displaymath}
{\mathbf \theta}_{1i} = \left[ \begin{array}{c}
\theta_{1i,1}\\
\theta_{1i,2}
\end{array} \right] =
\left[ \begin{array}{c}
\alpha_{1i}(\tau_{0}-\bar{\Delta}_{1i})\\
\alpha_{1i}
\end{array} \right]
\end{displaymath}
with ${\mathbf H}$ as in (\ref{eq:H}) and ${\mathbf W}_{1i} =
[W_{1i,1},\dots, W_{1i,m}]^{T}$. Since $\Psi_{1i,l+1}$ is an
independent Gaussian random variable for each $l$, ${\mathbf W}_{1i}
\sim {\mathcal N}(0,\Sigma_{1i})$ with $\Sigma_{1i} =
\sigma^{2}{\mathbf I}_{m}$. As mentioned, this set of observations
is for any node $1i$ in the set of $R_{1}$ nodes.

Since we have $\bar{N}$ $R_{1}$ nodes, we can write the vector of
observations made by all $R_{1}$ nodes as
\begin{equation} \label{eq:R1bigYvec}
\bar{{\mathbf Y}}_{1} = \bar{{\mathbf H}}\bar{{\mathbf \theta}}_{1}
+ \bar{{\mathbf W}}_{1}
\end{equation}
where
\begin{eqnarray*}
\bar{{\mathbf Y}}_{1} = \left[ \begin{array}{c} {\mathbf Y}_{11} \\
\vdots \\
{\mathbf Y}_{1\bar{N}}
\end{array} \right], \quad \bar{{\mathbf \theta}}_{1} = \left[ \begin{array}{c} {\mathbf \theta}_{11} \\
\vdots \\
{\mathbf \theta}_{1\bar{N}}
\end{array} \right], \quad \bar{{\mathbf W}}_{1}=\left[ \begin{array}{c} {\mathbf W}_{11} \\
\vdots \\
{\mathbf W}_{1\bar{N}}
\end{array} \right]
\end{eqnarray*}
and $\bar{{\mathbf H}}$ is as in (\ref{eq:barH}). Note that
$\bar{{\mathbf W}}_{1}\sim {\mathcal N}(0,\sigma^{2}{\mathbf
I}_{\bar{N}m})$. This way we have $\bar{{\mathbf Y}}_{1}$ as the
vector of observations made by all $R_{1}$ nodes and we can make a
UMVU (uniformly minimum variance unbiased) estimate of
$\bar{{\mathbf \theta}}_{1}$ by taking
\begin{eqnarray*}
\hat{\bar{{\mathbf \theta}}}_{1} =
(\bar{\mathbf{H}}^{T}\bar{\mathbf{H}})^{-1}\bar{\mathbf{H}}^{T}\bar{{\mathbf
Y}}_{1} \sim {\mathcal N}\left(\bar{\mu}_{1},\bar{\Sigma}_{1}
\right),
\end{eqnarray*}
where
\begin{equation*} 
\bar{\mu}_{1} = \bar{{\mathbf \theta}}_{1}, \quad \bar{\Sigma}_{1} =
\sigma^{2}(\bar{\mathbf{H}}^{T}\bar{\mathbf{H}})^{-1}.
\end{equation*}
It is easy to see that
\begin{eqnarray*}  
\lefteqn{(\bar{\mathbf{H}}^{T}\bar{\mathbf{H}})^{-1}} \nonumber \\
&& = \left[
\begin{array}{ccc} (\mathbf{H}^{T}\mathbf{H})^{-1}  &
\ldots & 0 \\
\vdots  & \ddots & \vdots \\
0 &  \ldots & (\mathbf{H}^{T}\mathbf{H})^{-1}
\end{array} \right]
\end{eqnarray*}
and
\begin{eqnarray*} 
(\mathbf{H}^{T}\mathbf{H})^{-1} = \left[ \begin{array}{cc}
\frac{2(2m-1)}{m(m+1)} & \frac{-6}{dm(m+1)} \\
\frac{-6}{dm(m+1)} & \frac{12}{d^2(m-1)m(m+1)}
\end{array} \right].
\end{eqnarray*}
This establishes the initial conditions for the theorem.
$\hat{\bar{{\mathbf \theta}}}_{1}$ is a $2\bar{N}\times 1$ column
vector where the subvector made up of the ($2(i-1)+1$)th and
($2(i-1)+2$)th elements, $i=1,\ldots,\bar{N}$, is $\hat{{\mathbf
\theta}}_{1i}=(\mathbf{H}^{T}\mathbf{H})^{-1}\mathbf{H}^{T}\mathbf{Y}_{1i}$.
Therefore, any node $1i$'s skew estimate (\ref{eq:alphahat}) and
offset estimate (\ref{eq:Deltahat}) can be found from
$\hat{\bar{{\mathbf \theta}}}_{1}$ as
\begin{equation} \label{eq:alphahat1i}
\hat{\alpha}_{1i} = e_{2(i-1)+2}^{T}\hat{\bar{{\mathbf \theta}}}_{1}
\end{equation}
and
\begin{equation} \label{eq:Deltahat1i}
\hat{\Delta}_{1i} = e_{2(i-1)+1}^{T}\hat{\bar{{\mathbf \theta}}}_{1}
- \tau_{0}
\end{equation}
where $e_{l}$ is the column vector of all zeros except for a one in
the $l$th position.

Each node $1i$ can now make an estimate of the next appropriate
integer value of $t$, in this case $t=\tau_{0}+md$, by making a
minimum variance unbiased estimate of
$\theta_{1i,1}+md\theta_{1i,2}=\alpha_{1i}(\tau_{0}-\bar{\Delta}_{1i})+md\alpha_{1i}$.
This can be done with the estimator
\begin{eqnarray*}
\hat{\tau}_{1i}=\hat{\theta}_{1i,1}+md\hat{\theta}_{1i,2} = {\mathbf
C}_{0}\hat{{\mathbf \theta}}_{1i}
\end{eqnarray*}
where ${\mathbf C}_{0}=[1 \quad md]$.  This will then be node $1i$'s
estimate of the next appropriate integer value of $t$ in its own
time scale $c_{1i}$.

From (\ref{eq:pulsetimeestimate}), every node $1i$ will then
transmit a sequence of $m$ pulses occurring, in the time scale of
$c_{1i}$, at $X_{1i}(l)=\hat{\tau}_{1i}+ld\hat{\theta}_{1i,2}$, for
$l=0,\ldots,m-1$. Using the clock model (\ref{eq:clock}), we find
that in the time scale of $c_{1}$ these pules occur at
\begin{eqnarray*}
(\hat{\tau}_{1i}+ld\hat{\theta}_{1i,2})_{c_{1}} &=&
\frac{\hat{\tau}_{1i}+ld\hat{\theta}_{1i,2}-\Psi_{1i,l+1}}{\alpha_{1i}}+\bar{\Delta}_{1i}
\\ &=& \frac{\hat{\tau}_{1i}}{\alpha_{1i}}+ \bar{\Delta}_{1i} +
ld\frac{\hat{\theta}_{1i,2}}{\alpha_{1i}}-
\frac{\Psi_{1i,l+1}}{\alpha_{1i}}.
\end{eqnarray*}
Any node $2j$ in the $R_{2}$ set of nodes that can hear node $1i$
will thus get a sequence of pulses
\begin{eqnarray*}
\lefteqn{\tilde{{\mathbf Y}}_{2j}[l+1]} \\ &=&
\alpha_{2j}\bigg(\bigg(\frac{\hat{\tau}_{1i}}{\alpha_{1i}}+
\bar{\Delta}_{1i} + ld\frac{\hat{\theta}_{1i,2}}{\alpha_{1i}}-
\frac{\Psi_{1i,l+1}}{\alpha_{1i}}\bigg)-\bar{\Delta}_{2j}\bigg)
\\
&& \hspace{2.5cm}+\Psi_{2j,l+1},
\end{eqnarray*}
where $l=0,\ldots,m-1$.

In this Type I network deployment every node $2j$ hears the same set
of $\bar{N}$ nodes and takes the sample mean of each cluster of
pulses for its observation, we can express the actual vector of
observations made by node $2j$ as
\begin{eqnarray*}
\lefteqn{{\mathbf Y}_{2j}[l+1]} \\
&=&\sum_{i=1}^{\bar{N}}
\frac{\alpha_{2j}}{\bar{N}}\bigg(\bigg(\frac{\hat{\tau}_{1i}}{\alpha_{1i}}+
\bar{\Delta}_{1i} + ld\frac{\hat{\theta}_{1i,2}}{\alpha_{1i}}-
\frac{\Psi_{1i,l+1}}{\alpha_{1i}}\bigg)-\bar{\Delta}_{2j}\bigg)\\
&& \hspace{2.5cm}+\Psi_{2j,l+1},
\end{eqnarray*}
where $l=0,\ldots,m-1$. Note that since these pulse arrivals are
clustered, we assume that for a given cluster, each pulse arrival is
corrupted by the same jitter.  Thus, receiver side jitter
$\Psi_{2j,l+1}$ is an independent sample for every $l$, but takes
the same value for each $i$.  This models the fact that clock errors
occurring in a small time window are highly correlated while errors
farther apart in time are independent.  We can rewrite this simply
as ${\mathbf Y}_{2j}[l+1] =
\alpha_{2j}((\tau_{1}+ld\tilde{\alpha}_{1}-\tilde{\Psi}_{1,l+1})-\bar{\Delta}_{2j})+\Psi_{2j,l+1}$,
where
\begin{eqnarray*}
\tau_{1} &\stackrel{\Delta}{=}&
\frac{1}{\bar{N}}\sum_{i=1}^{\bar{N}}\frac{\hat{\tau}_{1i}}{\alpha_{1i}}+
\bar{\Delta}_{1i} \qquad \tilde{\alpha}_{1} \stackrel{\Delta}{=}
\frac{1}{\bar{N}}\sum_{i=1}^{\bar{N}}\frac{\hat{\theta}_{1i,2}}{\alpha_{1i}}
\\
\tilde{\Psi}_{1,l+1} &\stackrel{\Delta}{=}&
\frac{1}{\bar{N}}\sum_{i=1}^{\bar{N}}\frac{\Psi_{1i,l+1}}{\alpha_{1i}}.
\end{eqnarray*}
Since every node $2j$ will see the same $\bar{N}$, this means that
every node $2j$ will have the same $\tau_{1}$ and
$\tilde{\alpha}_{1}$. Therefore, $\tau_{1}$ and $\tilde{\alpha}_{1}$
are now fixed, and it can be easily found that
\begin{displaymath}
\left[ \begin{array}{c} \tilde{\Psi}_{1,1} \\ \vdots \\
\tilde{\Psi}_{1,m}
\end{array}\right] \sim {\mathcal
N}\bigg(0,\Sigma_{\tilde{\Psi}_{1}}\bigg)
\end{displaymath}
where
\begin{displaymath}
\Sigma_{\tilde{\Psi}_{1}} =
\frac{\sigma^{2}}{\bar{N}^{2}}\sum_{i=1}^{\bar{N}}\frac{1}{\alpha_{1i}^{2}}{\mathbf
I}_{m}.
\end{displaymath}

Node $2j$'s vector of observations can also be written in a linear
form similar to (\ref{eq:linearFormCoop}), ${\mathbf
Y}_{2j}={\mathbf H}\theta_{2j} + {\mathbf W}_{2j}$, where
\begin{displaymath}
{\mathbf \theta}_{2j} = \left[ \begin{array}{c}
\theta_{2j,1}\\
\theta_{2j,2}
\end{array} \right] =
\left[ \begin{array}{c}
\alpha_{2j}(\tau_{1}-\bar{\Delta}_{2j})\\
\alpha_{2j}\tilde{\alpha}_{1}
\end{array} \right]
\end{displaymath}
with ${\mathbf H}$ as in (\ref{eq:H}) and ${\mathbf W}_{2j} =
[W_{2j,1}\dots W_{2j,m}]^{T}$.
\begin{displaymath}
{\mathbf W}_{2j} = \alpha_{2j}\left[ \begin{array}{c}
\tilde{\Psi}_{1,1}
\\ \vdots \\ \tilde{\Psi}_{1,m}
\end{array}\right]+\left[ \begin{array}{c} \Psi_{2j,1} \\ \vdots \\ \Psi_{2j,m}
\end{array}\right]\sim {\mathcal N}(0,\Sigma_{2j})
\end{displaymath}
with
\begin{displaymath}
\Sigma_{2j} =
\sigma^{2}\big(1+\frac{\alpha_{2j}^{2}}{\bar{N}^{2}}\sum_{i=1}^{\bar{N}}\frac{1}{\alpha_{1i}^{2}}\big){\mathbf
I}_{m}.
\end{displaymath}
The vector of observations made by all $R_{2}$ nodes can be written
in a manner similar to (\ref{eq:R1bigYvec}),
\begin{equation*} 
\bar{{\mathbf Y}}_{2} = \bar{{\mathbf H}}\bar{{\mathbf \theta}}_{2}
+ \bar{{\mathbf W}}_{2}
\end{equation*}
where
\begin{eqnarray*}
\bar{{\mathbf Y}}_{2} = \left[ \begin{array}{c} {\mathbf Y}_{21} \\
\vdots \\
{\mathbf Y}_{2\bar{N}}
\end{array} \right], \quad \bar{{\mathbf \theta}}_{2} = \left[ \begin{array}{c} {\mathbf \theta}_{21} \\
\vdots \\
{\mathbf \theta}_{2\bar{N}}
\end{array} \right], \quad {\mathbf Q}_{2} = \left[ \begin{array}{c} \alpha_{21}{\mathbf I}_{m} \\
\vdots \\
\alpha_{2\bar{N}}{\mathbf I}_{m}
\end{array} \right]
\end{eqnarray*}
\begin{displaymath}
\bar{{\mathbf W}}_{2} = \left[ \begin{array}{c} {\mathbf W}_{21} \\
{\mathbf W}_{22} \\
\vdots \\
{\mathbf W}_{2\bar{N}}
\end{array} \right]={\mathbf Q}_{2}\left[ \begin{array}{c}
\tilde{\Psi}_{1,1}
\\ \vdots \\ \tilde{\Psi}_{1,m}
\end{array}\right]+\left[ \begin{array}{c} \Psi_{21,1} \\ \vdots \\
\Psi_{21,m} \\ \vdots \\ \Psi_{2\bar{N},1} \\ \vdots \\
\Psi_{2\bar{N},m}
\end{array}\right]
\end{displaymath}
This means that $\bar{{\mathbf W}}_{2}\sim{\mathcal
N}(0,\Sigma_{\bar{{\mathbf W}}_{2}})$, where
\begin{eqnarray*}
\Sigma_{\bar{{\mathbf W}}_{2}} = {\mathbf
Q}_{2}\Sigma_{\tilde{\Psi}_{1}}{\mathbf Q}_{2}^{T} + \sigma^2
{\mathbf I}_{\bar{N}m}.
\end{eqnarray*}

The $R_{2}$ nodes will estimate $\bar{{\mathbf \theta}}_{2}$ as
\begin{eqnarray} \label{eq:theta2barhat}
\hat{\bar{{\mathbf \theta}}}_{2} &=&
(\bar{\mathbf{H}}^{T}\bar{\mathbf{H}})^{-1}\bar{\mathbf{H}}^{T}\bar{{\mathbf
Y}}_{2} \\
&\sim& {\mathcal
N}\left(\bar{\mathbf{\theta}}_{2},(\bar{\mathbf{H}}^{T}\bar{\mathbf{H}})^{-1}\bar{\mathbf{H}}^{T}\Sigma_{\bar{{\mathbf
W}}_{2}}((\bar{\mathbf{H}}^{T}\bar{\mathbf{H}})^{-1}\bar{\mathbf{H}}^{T})^{T}
\right) \nonumber .
\end{eqnarray}
However, for analysis, this does not give us the complete
distribution of $\hat{\bar{{\mathbf \theta}}}_{2}$ since
$\bar{\mathbf{\theta}}_{2}$ is a function of
$\hat{\bar{\mathbf{\theta}}}_{1}$.  Therefore, we first consider how
$\mathbf{\theta}_{2j}$ is a function of
$\hat{\bar{\mathbf{\theta}}}_{1}$.  We find that
\begin{eqnarray} \label{eq:theta2iasfxnoftheta1hat}
\mathbf{\theta}_{2j} & = & \left[ \begin{array}{c}
\alpha_{2j}(\tau_{1}-\bar{\Delta}_{2j})\\
\alpha_{2j}\tilde{\alpha}_{1}
\end{array} \right] \nonumber \\
& = & \left[ \begin{array}{c}
\alpha_{2j}(\frac{1}{\bar{N}}\sum_{i=1}^{\bar{N}}\frac{\hat{\tau}_{1i}}{\alpha_{1i}}+
\bar{\Delta}_{1i}-\bar{\Delta}_{2j})\\
\alpha_{2j}\frac{1}{\bar{N}}\sum_{i=1}^{\bar{N}}\frac{\hat{\theta}_{1i,2}}{\alpha_{1i}}
\end{array} \right] \nonumber \\
& = & \left[ \begin{array}{c}
\alpha_{2j}(\frac{1}{\bar{N}}\sum_{i=1}^{\bar{N}}\frac{\hat{\theta}_{1i,1}+md\hat{\theta}_{1i,2}}{\alpha_{1i}}+
\bar{\Delta}_{1i}-\bar{\Delta}_{2j})\\
\alpha_{2j}\frac{1}{\bar{N}}\sum_{i=1}^{\bar{N}}\frac{\hat{\theta}_{1i,2}}{\alpha_{1i}}
\end{array} \right] \nonumber \\
& = & \frac{\alpha_{2j}}{\bar{N}}\sum_{i=1}^{\bar{N}}\bigg(\left[
\begin{array}{cc} \frac{1}{\alpha_{1i}} & \frac{dm}{\alpha_{1i}} \\
0 & \frac{1}{\alpha_{1i}} \end{array} \right] \left[
\begin{array}{c} \hat{\theta}_{1i,1} \\ \hat{\theta}_{1i,2} \end{array} \right] \nonumber \\
&& \hspace{1.5cm} + \left[ \begin{array}{c} \bar{\Delta}_{1i} \\ 0
\end{array} \right] \bigg) - \alpha_{2j}\left[ \begin{array}{c}
\bar{\Delta}_{2j} \\ 0
\end{array} \right]
\end{eqnarray}
Using (\ref{eq:theta2iasfxnoftheta1hat}) we have
\begin{equation} \label{eq:theta2asfxnoftheta1hat}
\bar{\mathbf{\theta}}_{2} =
\mathbf{A}_{2}\hat{\bar{\mathbf{\theta}}}_{1} + \mathbf{B}_{2},
\end{equation}
where
\begin{eqnarray*}
\mathbf{A}_{2} = \mathbf{D}_{2}\left[ \begin{array}{ccccc}
\frac{1}{\alpha_{11}} & \frac{dm}{\alpha_{11}} & \ldots & 0
& 0 \\
0 & \frac{1}{\alpha_{11}}  & \ldots & 0 & 0 \\
\vdots & \vdots & \ddots & \vdots & \vdots \\
0 & 0 & \ldots & \frac{1}{\alpha_{1\bar{N}}} &
\frac{dm}{\alpha_{1\bar{N}}} \\
0 & 0 & \ldots & 0 & \frac{1}{\alpha_{1\bar{N}}}
\end{array}\right],
\end{eqnarray*}
\begin{eqnarray*}
\mathbf{B}_{2} = \mathbf{D}_{2}\left[ \begin{array}{c}
\bar{\Delta}_{11} \\ 0  \\ \vdots \\
\bar{\Delta}_{1\bar{N}} \\ 0
\end{array}\right] - \left[ \begin{array}{c}
\alpha_{21}\bar{\Delta}_{21} \\ 0 \\ \vdots \\
\alpha_{2\bar{N}}\bar{\Delta}_{2\bar{N}} \\ 0
\end{array}\right],
\end{eqnarray*}
for
\begin{eqnarray*}
\mathbf{D}_{2} = \frac{1}{\bar{N}}\left[ \begin{array}{ccccc}
\alpha_{21} & 0 &  \ldots & \alpha_{21}
& 0 \\
0 & \alpha_{21} & \ldots & 0 & \alpha_{21} \\
\vdots & \vdots & \ddots & \vdots & \vdots \\
\alpha_{2\bar{N}} & 0  & \ldots & \alpha_{2\bar{N}}
& 0 \\
0 & \alpha_{2\bar{N}} & \ldots & 0 & \alpha_{2\bar{N}}
\end{array}\right].
\end{eqnarray*}

Using (\ref{eq:theta2barhat}) and (\ref{eq:theta2asfxnoftheta1hat}),
the distribution of $\hat{\bar{{\mathbf \theta}}}_{2}$ can now be
found.
\begin{eqnarray} \label{eq:mu2}
\bar{\mu}_{2} &=& E(\hat{\bar{{\mathbf \theta}}}_{2}) \nonumber \\
&=& E(E(\hat{\bar{{\mathbf \theta}}}_{2}|\hat{\bar{{\mathbf
\theta}}}_{1})) \nonumber \\
&=& E(\bar{{\mathbf \theta}}_{2}) \nonumber \\
&=& E(\mathbf{A}_{2}\hat{\bar{\mathbf{\theta}}}_{1} +
\mathbf{B}_{2}) \nonumber \\
&=& \left[ \begin{array}{c}
\alpha_{21}(\tau_{0}+md-\bar{\Delta}_{21}) \\
\alpha_{21} \\
\vdots \\
\alpha_{2\bar{N}}(\tau_{0}+md-\bar{\Delta}_{2\bar{N}}) \\
\alpha_{2\bar{N}} \\
\end{array} \right]
\end{eqnarray}
Using the decomposition
\begin{displaymath}
\textrm{Cov}(\hat{\bar{{\mathbf
\theta}}}_{2})=E(\textrm{Cov}(\hat{\bar{{\mathbf
\theta}}}_{2}|\hat{\bar{{\mathbf
\theta}}}_{1}))+\textrm{Cov}(E(\hat{\bar{{\mathbf
\theta}}}_{2}|\hat{\bar{{\mathbf \theta}}}_{1})),
\end{displaymath}
we have from (\ref{eq:theta2barhat}) and
(\ref{eq:theta2asfxnoftheta1hat})
\begin{eqnarray*}
\Sigma_{m_{2}} & = & E(\textrm{Cov}(\hat{\bar{{\mathbf
\theta}}}_{2}|\hat{\bar{{\mathbf \theta}}}_{1})) \\
& = &
(\bar{\mathbf{H}}^{T}\bar{\mathbf{H}})^{-1}\bar{\mathbf{H}}^{T}\Sigma_{\bar{{\mathbf
W}}_{2}}((\bar{\mathbf{H}}^{T}\bar{\mathbf{H}})^{-1}\bar{\mathbf{H}}^{T})^{T}
\end{eqnarray*}
\begin{eqnarray*}
\textrm{Cov}(E(\hat{\bar{{\mathbf \theta}}}_{2}|\hat{\bar{{\mathbf
\theta}}}_{1})) & = & \textrm{Cov}(\bar{\theta}_{2}) \\
& = & \mathbf{A}_{2}\bar{\Sigma}_{1}\mathbf{A}_{2}^{T},
\end{eqnarray*}
giving us
\begin{equation} \label{eq:sigma2}
\bar{\Sigma}_{2} = \textrm{Cov}(\hat{\bar{{\mathbf \theta}}}_{2}) =
\Sigma_{m_{2}}+\mathbf{A}_{2}\bar{\Sigma}_{1}\mathbf{A}_{2}^{T}.
\end{equation}
Thus, the distribution of $\hat{\bar{{\mathbf \theta}}}_{2}$ is
\begin{equation*} 
\hat{\bar{{\mathbf \theta}}}_{2} \sim {\mathcal N}(\bar{\mu}_{2},
\bar{\Sigma}_{2}).
\end{equation*}
$\hat{\bar{{\mathbf \theta}}}_{2}$ is again a $2\bar{N}\times 1$
column vector where the subvector made up of the ($2(i-1)+1$)th and
($2(i-1)+2$)th elements, $i=1,\ldots,\bar{N}$, is $\hat{{\mathbf
\theta}}_{2i}=(\mathbf{H}^{T}\mathbf{H})^{-1}\mathbf{H}^{T}\mathbf{Y}_{2i}$.
Therefore, as in (\ref{eq:alphahat1i}) and (\ref{eq:Deltahat1i}),
any node $2i$'s skew estimate (\ref{eq:alphahat}) and offset
estimate (\ref{eq:Deltahat}) can be found from $\hat{\bar{{\mathbf
\theta}}}_{2}$ as
\begin{equation} \label{eq:alphahat2i}
\hat{\alpha}_{2i} = e_{2(i-1)+2}^{T}\hat{\bar{{\mathbf \theta}}}_{2}
\end{equation}
and
\begin{equation} \label{eq:Deltahat2i}
\hat{\Delta}_{2i} = e_{2(i-1)+1}^{T}\hat{\bar{{\mathbf \theta}}}_{2}
- (\tau_{0}+dm).
\end{equation}

Each node $2i$ will now be able to transmit a sequence of $m$ pulses
occurring, in the time scale of $c_{2i}$, at
$X_{2i}(l)=\hat{\tau}_{2i}+ld\hat{\theta}_{2i,2}$, for
$l=0,\ldots,m-1$, where
$\hat{\tau}_{2i}=\hat{\theta}_{2i,1}+md\hat{\theta}_{2i,2}$.
Repeating the same process we carried out for the observations of
any node $2j$ with any node $3j$, we can find $\hat{\bar{{\mathbf
\theta}}}_{3} \sim {\mathcal N}(\bar{\mu}_{3}, \bar{\Sigma}_{3})$.
In fact, continuing this procedure, we can find the distribution of
$\hat{\bar{{\mathbf \theta}}}_{k}$ for the $R_{k}$ nodes as
\begin{equation*} 
\hat{\bar{{\mathbf \theta}}}_{k} \sim {\mathcal N}(\bar{\mu}_{k},
\bar{\Sigma}_{k})
\end{equation*}
where similar to (\ref{eq:mu2}) we have $\bar{\mu}_{k} =
E(\hat{\bar{{\mathbf \theta}}}_{k})$ which is found in
(\ref{eq:muk}) and similar to (\ref{eq:sigma2}) we have
\begin{equation*} 
\bar{\Sigma}_{k} = \textrm{Cov}(\hat{\bar{{\mathbf \theta}}}_{k}) =
\Sigma_{m_{k}}+\mathbf{A}_{k}\bar{\Sigma}_{k-1}\mathbf{A}_{k}^{T},
\end{equation*}
which is found in (\ref{eq:sigmak}). $\Sigma_{m_{k}}$,
$\mathbf{A}_{k}$, $\mathbf{B}_{k}$, and $\Sigma_{\bar{{\mathbf
W}}_{k}}$ are as in the theorem statement. As in
(\ref{eq:alphahat2i}) and (\ref{eq:Deltahat2i}), any node $ki$'s
skew estimate (\ref{eq:alphahat}) and offset estimate
(\ref{eq:Deltahat}) can be found from $\hat{\bar{{\mathbf
\theta}}}_{k}$ as
\begin{equation*}
\hat{\alpha}_{ki} = e_{2(i-1)+2}^{T}\hat{\bar{{\mathbf \theta}}}_{k}
\end{equation*}
and
\begin{equation*}
\hat{\Delta}_{ki} = e_{2(i-1)+1}^{T}\hat{\bar{{\mathbf \theta}}}_{k}
- (\tau_{0}+dm(k-1)).
\end{equation*}
This concludes the proof of Theorem~\ref{theorem:main}. \qquad
$\bigtriangleup$

\end{document}